\documentstyle[prb,aps,floats,psfig]{revtex}

\newcommand{\gmathbf}[1]{{\mathbf #1}}
\newcommand{\gmathrm}[1]{{\mathrm #1}}
\newcommand{\gmathcal}[1]{{\mathcal #1}}

\begin{document}

\title{{\bf Spin-wave spectrum of a two-dimensional itinerant electron
    system:\\
Analytic results for the incommensurate spiral phase in the strong-coupling limit}}
\par
\author{E. Arrigoni \\ Institut f\"{u}r Theoretische Physik, Universit\"{a}t
W\"{u}rzburg \\ D-97074 W\"{u}rzburg, Germany \\
G.C. Strinati \\ Dipartimento di Matematica e Fisica, 
Sezione INFM \\ Universit\`{a} di Camerino, I-62032 Camerino (MC), Italy}

\date{\today}
\maketitle
\begin{abstract}
We study the zero-temperature spin fluctuations of a two-dimensional 
itinerant-electron system with an incommensurate magnetic ground state described 
by a single-band Hubbard Hamiltonian.
We introduce the (broken-symmetry) magnetic phase at the mean-field (Hartree-Fock) 
level through a \emph{spiral spin configuration\/} with characteristic wave vector 
$\gmathbf{Q}$ different in general from the antiferromagnetic wave vector 
$\gmathbf{Q_{AF}}$, and consider spin fluctuations over and above it within the 
electronic random-phase (RPA) approximation.
We obtain a \emph{closed\/} system of equations for the generalized wave vector
and frequency dependent susceptibilities,  which are equivalent to the ones
reported recently by Brenig.
We obtain, in addition, analytic results for the spin-wave dispersion relation in 
the strong-coupling limit of the Hubbard Hamiltonian and find that at finite doping 
the spin-wave dispersion relation has a \emph{hybrid form\/} between that associated 
with the (localized) Heisenberg model and that associated with the (long-range) RKKY 
exchange interaction.
We also find an instability of the spin-wave spectrum in a finite region about the 
center of the Brillouin zone, which signals a physical instability toward a different 
spin- or, possibly, charge-ordered phase, as, for example, the  stripe structures observed 
in the high-Tc materials. 
We expect, however, on physical grounds that for wave vectors external to this 
region the spin-wave spectrum that we have determined should survive consideration 
of more sophisticated mean-field solutions. 
\end{abstract}


\section{Introduction}

Spin fluctuations about incommensurate static magnetic configurations represent an 
interesting problem, especially in the view of the accumulating experimental evidence 
on the parent compounds of the high-$T_{c}$ cuprate materials.\cite{spiral-data}
The dynamic structure factor $S(\gmathbf{q},\omega)$ at wave vector $\gmathbf{q}$
and frequency $\omega$, as measured by inelastic neutron scattering, shows noticeable
peaks over the background when $\gmathbf{q}=\gmathbf{Q}=\gmathbf{Q_{AF}} +
\Delta~\gmathbf{Q}$ ($|\Delta~\gmathbf{Q}| \ll |\gmathbf{Q_{AF}}|$) for increasing
dopant concentration, at finite albeit small values of $\omega$ (with $\omega \ll J$,
$J$ being the exchange coupling).\cite{finite-omega}
Further experimental studies for larger values of $|\gmathbf{q}-\gmathbf{Q_{AF}}|$
have also detected the presence of well-defined spin-wave-type excitations close
to the boundary of the antiferromagnetic (AF) Brillouin zone (BZ).\cite{Aeppli}
The fact that spin-wave excitations with small wavelength (i.e., comparable to
lattice spacing) can be detected even in the absence of long-range magnetic order,
has actually been well established experimentally since the eighties also in more
conventional magnetic materials.\cite{Mook}
This state of affairs has, in turn, prompted a number of theoretical studies on 
dynamical excitations about incommensurate spin 
configurations.\cite{Shulz,SS,Brenig,Fulde}

From the theoretical point of view, inclusion of incommensurate spin configurations 
in cluster mean-field calculations would require one to consider cluster sizes at least 
as big as the spatial extension of the incommensurability.
In this way, complex incommensurate patterns could as well be included in the 
calculation, at the price of considerable numerical effort and without full analytic 
control on the results.
Alternatively, one may set up calculations for an infinite system with necessarily
simpler incommensurate spin configurations, with the advantage, however, of
obtaining analytic results (at least) in some limits, which in turn may admit a
simple physical interpretation.
In this respect, the \emph{spiral spin configuration\/} appears to be the only 
one for which \emph{analytic\/} calculations can be performed, in the sense that 
the constituent equations can be brought to a \emph{closed form\/} which is
manageable for controlled analytic approximations.
Indeed previous analytic calculations have considered \emph{static\/} long-range 
background spiral spin configurations, on top of which dynamical spin excitations 
have been considered.\cite{Shulz,Brenig}

The intrinsic simplicity of the background spiral spin configuration and the neglect 
of its dynamics have resulted, however, into an instability of the spin-wave spectrum 
in a limited region of the Brillouin zone.\cite{Brenig}
This outcome could, in principle, make the resulting spin-wave spectrum altogether 
unreliable, as the underlying spin configuration (over and above which spin-wave
excitations are constructed) could be much more complex 
than a spiral one and
even possess a dynamics of its own.
In addition, a static or dynamic charge modulation can be present as well, like the stripe 
structures observed in some underdoped cuprates.\cite{tr.st.95} 
One expects, however, on physical grounds the \emph{short-wavelength\/} spin-wave 
excitations, obtained on top of a spiral configuration, to preserve their dispersion 
relation even when considering more complex long-range underlying structures.

Within this framework, we have pursued the analytic calculation with an underlying
\emph{spiral\/} spin configuration for an itinerant electronic system described by
a single-band Hubbard Hamiltonian.
The spin (and charge) dynamics have bee described within the electronic random-phase
(RPA) approximation for the dynamical susceptibilities, based on a broken-symmetry 
Hartree-Fock (HF) mean-field solution, along the lines of Ref.~\cite{Brenig}.
In particular, we have considered the large $U/t$ limit in detail (where $t$ is
the nearest-neighbor hopping matrix element and $U$ is the local on-site repulsion
in the Hubbard Hamiltonian), for which the spin-wave dispersion relation (as obtained
from the poles of the dynamical susceptibilities) can be expressed in analytic
form by systematically expanding in the \emph{small parameter\/} $t/U$.
In this way, we have obtained an analytic expression for the spin-wave dispersion 
relation valid in the limit $t/U \ll 1$ [cf. Eq.~(\ref{disp-rel-4}) below],  showing a 
novel characteristic structure, 
namely, a hybrid form between the dispersion relations obtained within the
nearest-neighbor Heisenberg model for localized spins \cite{Keffer} and the long-range
RKKY interaction mediated by the conduction electrons.\cite{RK}

This analytic form of the dispersion relation could admittedly not have been guessed
\emph{a priori\/}, by fitting the dispersion relation obtained numerically (from the 
location of the poles of the dynamical susceptibilities) with a Heisenberg model 
extending in principle to a large albeit \emph{finite\/} number of neighbors.
Neither, this analytic form can be simply reduced to a nearest-neighbor Heisenberg
dispersion relation with a doping-dependent exchange integral.\cite{tJmodel}
Rather, the characteristic long-range RKKY contribution would require fitting to
a Heisenberg model with an infinite number of neighbors.
This would contradict the spirit with which the Heisenberg model was
introduced to start with,\cite{Heisenberg} namely, as a fitting model that makes
physical sense when the interactions extend to a limited number of neighbors only.
Note that this situation contrasts that found at half-filling of the
Hubbard band (i.e., in the absence of doping), where the antiferromagnetic (AF)
spin-wave spectrum can be nicely fitted by a Heisenberg model extending at most 
to a few neighbors.\cite{AS-92}
Our results also show that the magnitude of the overall exchange integral, which
characterizes the spin-wave spectrum, decreases with increasing doping and vanishes
when the transition to a ferromagnetic case occurs at the mean-field level.

The plan of the paper is as follows. 
Section 2 obtains the implicit form of the spin-wave dispersion relation within 
the HF-RPA approximations, by solving for the dynamical susceptibilities of the 
itinerant electron system in the presence of an incommensurate spin spiral ground 
state.
Section 3 focuses on the small $t/U$ expansion of the results of Section 2, which 
requires a careful analysis of the doping dependence of the relevant quantities.
Section 4 discusses the main results of this paper and Section 5 gives our
conclusions.
For the sake of completeness, we report in the Appendices  details of the analytic
calculations, and adapt know results for the Heisenberg and RKKY spin-wave spectra
to the present context.


\vspace{2cm}
\section{Dynamical susceptibilities within the itinerant-electron RPA approximation 
with an incommensurate spin-spiral ground state}

In this Section, we give the derivation of the spin-wave dispersion relation
for a two-dimensional itinerant-electron system in the presence of an incommensurate 
spiral magnetic structure \emph{with a generic characteristic wave vector\/} $\gmathbf{Q}$, 
within the electronic RPA approximation.
Although our results for the dispersion relation coincide with those previously 
given by Brenig in Ref.~\cite{Brenig}, we provide here in addition the expression of the
correlation functions which can be relevant for a direct comparison with neutron
scattering experiments.
Further details of the calculation are reported in Appendix A.

We emphasize that the finding of a closed-form expression for the correlation functions 
for \emph{any\/} characteristic wave vector $\gmathbf{Q}$ (incommensurate with the lattice
spacing) and not just for the (commensurate) antiferromagnetic wave vector $\gmathbf{Q}_{AF}$ 
is altogether a nontrivial result, being intrinsically related to the peculiar pattern of 
the spiral magnetic solution for the ground state.

We begin by considering the generalized \emph{correlation function\/}
at zero temperature in the broken-symmetry phase:

\begin{equation}
\gmathcal{X}_{\mu,\nu}(\gmathbf{r} t ,\gmathbf{r'} t') = -i \left<T[S_{\mu}(\gmathbf{r},t)
       \, S_{\nu}(\gmathbf{r'},t')]\right> + i \left<S_{\mu}(\gmathbf{r},t)\right> 
       \left<S_{\nu}(\gmathbf{r'},t')\right>                         \label{susceptibilities}   
\end{equation}

\noindent
where the average $\left<\cdots\right>$ is taken over the ground state, $T$ stands here
for Wick's time-ordering operator, and $S_{\mu}(\gmathbf{r})$ is given by
(we set $\hbar = 1$ throughout)

\begin{equation}
S_{\mu}(\gmathbf{r}) = \frac{1}{2} \sum_{\alpha, \beta} \, 
\psi^{\dagger}_{\alpha}(\gmathbf{r}) \, \sigma^{\mu}_{\alpha,\beta} \,
                 \psi_{\beta}(\gmathbf{r})  \,\, .                    \label{spin-operator-0}         
\end{equation}

\noindent
In these expressions, $\mu,\nu = (0,x,y,z)$, $\sigma^{\mu}$ is a Pauli matrix
(with $\sigma^{0}$ equal to the $2\times2$ identity matrix), and $\alpha, \beta$ 
are spin labels.
Note that $2 S_{0}(\gmathbf{r})$ coincides with the density operator, which couples 
with the spin operator in the presence of an incommensurate spiral magnetic structure.

For the simple band we are considering, the field operator in Eq.(\ref{spin-operator-0})
can be represented in the form

\begin{equation} 
\psi_{\alpha}(\gmathbf{r}) = \sum_{i}\, \phi(\gmathbf{r}-\gmathbf{R}_{i}) \,
                            c_{i \alpha}   \,\,  ,                    \label{field-operator}                           
\end{equation}

\noindent      
where $\phi(\gmathbf{r})$ is the atomic (Wannier) orbital associated with the
simple band, $\gmathbf{R}_{i}$ is the lattice vector locating site $i$, 
and $c_{i \alpha}$ is a destruction operator. 
Time evolution in Eq.~(\ref{susceptibilities}) is governed by the Heisenberg 
representation:

\begin{equation}
\psi_{\alpha}(\gmathbf{r},t) = \gmathrm{e}^{iHt} \, \psi_{\alpha}(\gmathbf{r}) 
                      \, \gmathrm{e}^{-iHt}                            \label{Heisenberg-rep}
\end{equation}

\noindent
where for $H$ we take the simple-band two-dimensional Hubbard Hamiltonian.

In terms of the two-particle correlation function $L$,\cite{BK} the
generalized correlation function (\ref{susceptibilities}) takes the form

 \begin{equation}
\gmathcal{X}_{\mu,\nu}(\gmathbf{r} t ,\gmathbf{r'} t') = 
          - \frac{i}{4} \, \sum_{\alpha,\beta}\sum_{\alpha',\beta'} \,
          \sigma^{\mu}_{\alpha,\beta} \,\, \sigma^{\nu}_{\alpha',\beta'} \,\,
          L(\gmathbf{r} t \beta,\gmathbf{r'} t' \beta'; 
      \gmathbf{r} t^{+} \alpha,\gmathbf{r'} t'^{+} \alpha')               \label{definition-L}
\end{equation}

\noindent                                  
with $t^{+} = t + \eta$ ($\eta=0^{+}$), where $L$ satisfies the Bethe-Salpeter
equation:

\begin{eqnarray}
L(1,2;1',2') & = & G(1,2') \, G(2,1')                             \label{Bethe-Salpeter}  \\
& + & \int\! d3 \, d4 \, d5 \, d6 \, G(1,3) \, G(4,1') \, \Xi(3,5;4,6)\, 
                                                 L(6,2;5,2')       \nonumber
\end{eqnarray}

\noindent
($1,2,\cdots$ signifying the set of space, spin, and time variables).
In the above expression,

\begin{equation}
G(1,2) = -i \left<T\left[\psi(1)\psi^{\dagger}(2)\right]\right>           \label{single-p-G}
\end{equation}

\noindent
is the single-particle Green's function and the kernel $\Xi$ represents an
effective two-particle interaction. 
In particular, within the RPA approximation we are adopting, the kernel $\Xi$ takes 
the form:

\begin{eqnarray}
 \Xi(3,5;4,6) & = & -i v_{0} \, \delta(3,4) \, \delta(5,6^{+}) \, \delta(x_{3},x_{6}) 
                \, \delta(\alpha_3,-\alpha_6)    \nonumber      \\   
              & + & i v_{0} \, \delta(3,6) \, \delta(4,5) \, \delta(x_{3}^{+},x_{4})   
                \,  \delta(\alpha_{3},-\alpha_{4})                         \label{csi-RPA}
\end{eqnarray}

\noindent
with the notation $x\equiv(\gmathbf{r},t)$ and where the constant $v_{0}$ can be related 
to the parameter $U$ of the Hubbard Hamiltonian as follows:
 
\begin{equation} 
U = v_{0}\int \! d\gmathbf{r} \,\, |\phi(\gmathbf{r})|^{4}  \,\, .             \label{U-vs-vO}
\end{equation}

Entering Eqs.(\ref{Bethe-Salpeter}) and (\ref{csi-RPA}) into Eq.(\ref{definition-L}), 
we obtain for the generalized correlation function within the RPA approximation:

\begin{eqnarray}
\gmathcal{X}_{\mu,\nu}(x ,x') & = & -\frac{i}{4}
           \sum_{\alpha,\beta}\sum_{\alpha',\beta'} \,
          \sigma^{\mu}_{\alpha,\beta}  \,\sigma^{\nu}_{\alpha',\beta'} \,
          G(x \beta,x' \alpha') \, G(x' \beta',x \alpha)                \nonumber   \\       
& + &\frac{v_{0}}{4} \, (-i)^{2} \sum_{\alpha,\beta} \sum_{\alpha',\beta'} \,
          \sigma^{\mu}_{\alpha,\beta} \, \sigma^{\nu}_{\alpha',\beta'} 
  \int \! d3 \, G(x \beta,x_3 \alpha_3) \, G(x_3 \alpha_3,x \alpha)     \nonumber   \\
& \times & L(x_3 \bar{\alpha}_3,x' \beta'; x_3^{+} \bar{\alpha_3},x'^{+} \alpha')
                                                                        \nonumber   \\ 
& - & \frac{v_{0}}{4} \, (-i)^{2} \sum_{\alpha,\beta} \sum_{\alpha',\beta'} \,
          \sigma^{\mu}_{\alpha,\beta} \, \sigma^{\nu}_{\alpha',\beta'} 
  \int \! d3 \, G(x \beta,x_3 \alpha_3) \, G(x_3 \bar{\alpha_3},x \alpha) \nonumber  \\
& \times & L(x_3 \alpha_3,x' \beta'; x_3^{+} \bar{\alpha_3},x'^{+} \alpha')  \label{B-S-RPA} 
\end{eqnarray}

\noindent
with $\bar{\alpha}=-\alpha$. 
This is apparently not a closed-form equation for $\gmathcal{X}$ itself.
By the manipulations reported in Appendix A, however, Eq.~(\ref{B-S-RPA}) can be cast 
in the form of a coupled set of equations for the matrix components of the correlation 
function, as follows:

\begin{equation}
\gmathcal{X}_{\mu,\nu}(x,x') = \gmathcal{X}^{(0)}_{\mu,\nu}(x,x') 
+ 2 \, v_{0} \sum_{\mu',\nu'} \int \! dx'' \, \gmathcal{X}^{(0)}_{\mu,\mu'}(x,x'') \, 
\epsilon_{\mu',\nu'} \, \gmathcal{X}_{\nu',\nu}(x'',x')           \label{integral-equation-0}
\end{equation}

\noindent
where we have introduced the tensor

\begin{equation}
\epsilon_{\mu,\nu} = \left(
\begin{array}{rrrr} 1 &  0 &  0 &  0 \\
                    0 & -1 &  0 &  0 \\
                    0 &  0 & -1 &  0 \\
                    0 &  0 &  0 & -1
\end{array} \right)  \,\, . 
\end{equation}       

To solve Eq.~(\ref{integral-equation-0}), the explicit form of the \emph{non-interacting 
counterpart\/} $\gmathcal{X}^{(0)}$ of $\gmathcal{X}$ is required.
To this end, the ground-state average in Eq.~(\ref{single-p-G}) is evaluated as shown in
Appendix A, in terms of the eigenvalues ($\epsilon_{r}$) and eigenvectors ($W_{\xi,r}$) 
of the mean-field Hubbard Hamiltonian (see also Section 3), yielding for the time Fourier 
transform of $\gmathcal{X}^{(0)}$ the expression

\begin{eqnarray} 
\gmathcal{X}^{(0)}_{\mu,\nu}(\gmathbf{r},\gmathbf{r'}; \omega) & = & 
      \frac{1}{4\gmathcal{N}^{2}} \sum_{i,j} \sum_{\gmathbf{k},\gmathbf{k'}}^{BZ} \, 
\gmathrm{e}^{i(\gmathbf{k}-\gmathbf{k'}) \cdot(\gmathbf{R}_{i}-\gmathbf{R}_{j})} \,
|\phi(\gmathbf{r} -\gmathbf{R}_{i})|^{2} \, |\phi(\gmathbf{r'} -\gmathbf{R}_{j})|^{2} \nonumber   \\
& \times &  \sum_{r,r'} \sum_{\mu',\nu'} \,
T_{\mu,\mu'}(\Omega_{i}) \, T_{\nu,\nu'}(\Omega_{j}) \,
 F^{\mu'}_{r',r}(\gmathbf{k'},\gmathbf{k}) \,
 F^{\nu'}_{r,r'}(\gmathbf{k},\gmathbf{k'})                         \nonumber   \\ 
& \times &    \gmathcal{F}_{r,r'}(\gmathbf{k},\gmathbf{k'},\omega)             \label{FT-chi-O}
\end{eqnarray}

\noindent
where $BZ$ stands for the two-dimensional Brillouin zone, $\gmathcal{N}$ is the number of 
lattice sites, and the quantities $T$, $F$, and $\gmathcal{F}$ are defined in Appendix A 
($\Omega_{i}$ standing for the angles defining the local spin quantization axis). 
We consider further the space Fourier transform

\begin{eqnarray} 
\gmathcal{X}_{ab}(\gmathbf{q},\gmathbf{q'};\omega) & = &
 \frac{1}{\gmathcal{N} V_{0}} \int \! d \gmathbf{r} \, d \gmathbf{r'} \,
 \gmathrm{e}^{-i\gmathbf{q} \cdot \gmathbf{r}} \,
\gmathcal{X}_{ab}(\gmathbf{r},\gmathbf{r'};\omega)  
 \gmathrm{e}^{i\gmathbf{q'} \cdot \gmathbf{r'}}                       \nonumber \\
& \equiv & \frac{1}{V_{0}} \, S(\gmathbf{q}) \, S^{*}(\gmathbf{q'}) \,
\hat{\gmathcal{X}}_{ab}(\gmathbf{q},\gmathbf{q'};\omega)                        \label{FT2-chi}
\end{eqnarray}

\noindent
where $V_{0}$ is the volume of the elementary crystal cell and

\begin{equation}
S(\gmathbf{q}) \equiv \int \! d\gmathbf{r} \, \gmathrm{e}^{-i \gmathbf{q} \gmathbf{r}} \,
|\phi(\gmathbf{r})|^{2}                                                   \label{form-factor}
\end{equation}

\noindent
is a form factor (which can be set equal to unity for all practical purposes).
Applying a suitable unitary transformation [cf. Eq.~(\ref{A-T-bar-equal})] which renders 
the matrix $T(\Omega_{i})$ of Eq.~(\ref{FT-chi-O}) diagonal, one gets for its lattice 
Fourier transform:

\begin{eqnarray}
\bar{T}_{ab}(\gmathbf{k}) & = & \frac{1}{\gmathcal{N}} \sum_{i} \, 
\gmathrm{e}^{i\gmathbf{k} \cdot \gmathbf{R}_{i}} \, \bar{T}_{ab}(\Omega_{i})  \nonumber \\ 
& = & \left(
\begin{array}{cccc}
\Delta(\gmathbf{k}) & 0 & 0 & 0 \\
0 & \Delta(\gmathbf{k}-\gmathbf{Q}) & 0 & 0 \\
0 & 0 & \Delta(\gmathbf{k}) & 0 \\
0 & 0 & 0 & \Delta(\gmathbf{k}+\gmathbf{Q})
\end{array}
\right)                                                                        \label{T-bar-matrix}
\end{eqnarray}

\noindent
$\Delta(\gmathbf{k})$ being the lattice Kronecker delta function.
In the new basis, we thus obtain for the non-interacting correlation function
the expression:

\begin{eqnarray}
\hat{\gmathcal{X}}^{(0)}_{ab}(\gmathbf{q},\gmathbf{q'};\omega) & = &
     \frac{1}{4\gmathcal{N}} \sum_{\gmathbf{k} \gmathbf{k'}}^{BZ} \sum_{a', b'} \sum_{r, r'} \,
     \bar{T}_{aa'}(\gmathbf{k}-\gmathbf{k'}-\gmathbf{q})  \,
     \bar{T}_{b'b}(\gmathbf{k'}-\gmathbf{k}+\gmathbf{q'})        \nonumber    \\
& \times & \bar{F}^{a'}_{r', r}(\gmathbf{k'},\gmathbf{k}) \,
\bar{F}^{b'}_{r, r'} (\gmathbf{k},\gmathbf{k'}) \,
\gmathcal{F}_{r,r'}(\gmathbf{k},\gmathbf{k'},\omega)                          \label{chi-o-cap}
\end{eqnarray}

\noindent
where the overbar denotes matrices transformed according to the above unitary 
transformation [cf. Eqs.~(\ref{A-Fbar-definition}) and (\ref{A-T-bar-equal})].
In this way, the integral equation (\ref{integral-equation-0}) reduces to the form:

\begin{eqnarray}
\hat{\gmathcal{X}}_{ab}(\gmathbf{q},\gmathbf{q'};\omega) & = & 
\hat{\gmathcal{X}}^{(0)}_{ab}(\gmathbf{q},\gmathbf{q'};\omega)          \nonumber  \\   
& + & 2 U \sum_{a' b'} \sum_{\gmathbf{q''}} \,
\hat{\gmathcal{X}}^{(0)}_{aa'}(\gmathbf{q},\gmathbf{q''};\omega) \, \bar{\epsilon}_{a' b'} \,
\hat{\gmathcal{X}}_{b' b}(\gmathbf{q''},\gmathbf{q'};\omega)            \label{matrix-equation}
\end{eqnarray}

\noindent
with $\bar{\epsilon}$ given by Eq.~(\ref{A-epsilon-bar}).
This equation can be solved by the methods of Appendix A, yielding the closed-form
expression:

\begin{eqnarray}
\hat{\gmathcal{X}}_{ab}(\gmathbf{q}+\gamma_{a}\gmathbf{Q},\gmathbf{q'};\omega) & = &
\sum_{a'} \, [\gmathbf{1} + X(\gmathbf{q},\omega)]_{aa'}^{-1}         \nonumber  \\ 
& \times & X^{(0)}_{a'b}(\gmathbf{q}+\gamma_{a'}\gmathbf{Q};\omega|\gmathbf{Q}) \,
\Delta(\gmathbf{q}-\gmathbf{q'}-\gamma_{b}\gmathbf{Q})                              \label{xxx}
\end{eqnarray}

\noindent
where $\gmathbf{1}$ is the $4\times4$ unit matrix, the matrix $X(\gmathbf{q},\omega)$ is
defined by Eq.~(\ref{A-big-matrix}), and with the notation $\gamma_{a}=0$ for $a=0,2$, 
$\gamma_{a}=-1$ for $a=1$, and $\gamma_{a}=1$ for $a=3$.
Although still expressed in the transformed basis, Eq.~(\ref{xxx}) is the desired 
expression for the Fourier transform of the generalized correlation function, which holds 
within the RPA approximation \emph{for any value of\/} $\gmathbf{Q}$.

The spin-wave \emph{dispersion relation\/} can eventually be obtained by searching
for the zeros of the inverse matrix on the right-hand side of Eq.~(\ref{xxx}), which is 
equivalent to imposing the condition:

\begin{equation}
\det [\gmathbf{1} + X(\gmathbf{q},\omega)] \, = \, 0  \, .                        \label{zero}
\end{equation}

\noindent 
It can be verified that the condition (\ref{zero}) can be mapped onto the result
reported in Ref.~\cite{Brenig}, where the dispersion relation has then been obtained 
numerically for chosen values of $\gmathbf{Q}$.
In the present paper we proceed instead to deriving the \emph{analytic\/} form of the 
dispersion relation for \emph{small values of the parameter\/} $t/U$ of the Hubbard 
Hamiltonian.

To this end, it is convenient to rewrite first the matrix $X$ in Eq.~(\ref{zero}) in a 
more conventional basis identified by the labels ($0,+,-,z$), with 
$\sigma^{\pm}=(\sigma^{x} \pm i \sigma^{y})/\sqrt{2}$.
The matrix $\gmathbf{1} + X(\gmathbf{q},\omega)$ is then transformed into:
   
\begin{equation}
M(\gmathbf{q},\omega) \, = \, 
\gmathbf{1} + 2 U 
\left(
\begin{array}{cccc}
-\gmathcal{X}_{0}^{0,0 }(\gmathbf{q},\omega) & \gmathcal{X}_{0}^{0,-}(\gmathbf{q},\omega) &
\gmathcal{X}_{0}^{0,+}(\gmathbf{q},\omega) & \gmathcal{X}_{0}^{0,z}(\gmathbf{q},\omega)  \\
-\gmathcal{X}_{0}^{+,0}(\gmathbf{q},\omega) & \gmathcal{X}_{0}^{+,-}(\gmathbf{q},\omega) &
\gmathcal{X}_{0}^{+,+}(\gmathbf{q},\omega) & \gmathcal{X}_{0}^{+,z}(\gmathbf{q},\omega) \\    
-\gmathcal{X}_{0}^{-,0}(\gmathbf{q},\omega) & \gmathcal{X}_{0}^{-,-}(\gmathbf{q},\omega) &
\gmathcal{X}_{0}^{-,+}(\gmathbf{q},\omega) & \gmathcal{X}_{0}^{-,z}(\gmathbf{q},\omega) \\   
-\gmathcal{X}_{0}^{z,0}(\gmathbf{q},\omega) & \gmathcal{X}_{0}^{z,-}(\gmathbf{q},\omega) &
\gmathcal{X}_{0}^{z,+}(\gmathbf{q},\omega) & \gmathcal{X}_{0}^{z,z}(\gmathbf{q},\omega) \\
\end{array} \right)                                                     \label{definition-M} 
\end{equation}

\noindent
where now

\begin{equation}
\gmathcal{X}_{0}^{\alpha,\beta}(\gmathbf{q},\omega) =
\frac{1}{4\gmathcal{N}} \sum_{r,r'} \sum_{\gmathbf{k}}^{BZ} \,  
F^{\alpha}_{r',r}(\gmathbf{k}-\gmathbf{q},\gmathbf{k}) \,  
F^{\beta}_{r,r'}(\gmathbf{k},\gmathbf{k}-\gmathbf{q}) \,
\gmathcal{F}_{r,r'}(\gmathbf{k},\gmathbf{k}-\gmathbf{q},\omega)                       \label{da}           
\end{equation}

\noindent
and
  
\begin{equation} 
F^{\mu}_{r,r'}(\gmathbf{k},\gmathbf{k'}) =
\sum_{\xi,\xi'} \, W^{\dagger}_{r,\xi}(\gmathbf{k}) \, \sigma^{\mu}_{\xi,\xi'} \,
W_{\xi',r'}(\gmathbf{k'})\                                                         \label{du}                      
\end{equation}

\noindent 
with $\sigma^{+}=~\sqrt{2}\left(\begin{array}{cc} 0 & 1 \\ 0 & 0 \end{array} \right)$
and  $\sigma^{-}=~\sqrt{2}\left(\begin{array}{rr} 0 & 0 \\ 1 & 0 \end{array} \right)$.
Note that two columns in the expression (\ref{definition-M}) appear interchanged with 
respect to the original order, owing to the presence in the final basis of the tensor

\begin{equation}
\tilde{\epsilon} = \left(
\begin{array}{rrrr}
1 &  0 &  0 &  0 \\
0 &  0 & -1 &  0 \\
0 & -1 &  0 &  0 \\
0 &  0 &  0 & -1
\end{array}
\right)                                                                   \label{new-tensor}
\end{equation}

\noindent
in the place of $\bar{\epsilon}$ given by Eq.~(\ref{A-epsilon-bar}).

We pass now to perform the small $t/U$ expansion of the spin-wave dispersion relation
obtained from the condition $\det M(\gmathbf{q},\omega) = 0$.


\vspace{2cm}
\section{Expansion in the small parameter $\lowercase{t}/U$}

In this Section, we obtain explicitly the spin wave dispersion relation
to \emph{second order\/} in the small parameter $t/U$, from the general 
condition $\det M(\gmathbf{q},\omega) = 0$ obtained in Section 2 within the RPA 
approximation for the zero-temperature broken-symmetry phase with a generic 
incommensurate wave vector $\gmathbf{Q}$.
To this end, we will preliminary expand the self-consistency parameters of the 
mean-field Hamiltonian, as well as the equations they satisfy, at the 
relevant order in $t/U$; we will then expand the matrix elements of the matrix 
$M(\gmathbf{q},\omega)$ defined by Eq.~(\ref{definition-M}) at the relevant order 
in $t/U$.

\vspace{0.8cm}
\begin{center}
\begin{large}
{\bf A. Mean-field equations}
\end{large}
\end{center}
\vspace{0.8cm}

The mean-field equations for a single-band Hubbard Hamiltonian in the 
presence of an incommensurate spiral spin structure have been discussed in
Ref.\cite{AS-91}.
Introducing a local set of spin quantization axis, with the $z$ axis 
transformed locally into the axis specified by the spherical angles
$\Omega_{i}\equiv(\theta_{i}=\gmathbf{Q}\cdot\gmathbf{R}_{i},\varphi_{i}=0)$
at site $i$, one transforms the destruction operators $c_{i \alpha}$ according 
to Eq.~(\ref{A-fermion-field}) and performs the (Hartree-Fock) mean-field 
decoupling of the Hubbard Hamiltonian, yielding

\begin{equation} 
H(\gmathbf{Q}) \, = \, \sum_{\gmathbf{k}}^{BZ} \sum_{\xi,\xi'} \, 
d^{\dagger}_{\gmathbf{k} \xi} \,\gmathcal{H}_{\xi,\xi'}(\gmathbf{k})  \,
d_{\gmathbf{k} \xi'} \, - \, U \gmathcal{N} \left( m^{2}_{1} \, - \, m^{2}_{2} \right)       
                                                                        \label{Hamiltonian1}
\end{equation}

\noindent
with $\xi,\xi'=(+,-)$ and where $\gmathcal{H}$ is the $2\times2$ matrix

\begin{equation}
\gmathcal{H}(\gmathbf{k}) \, = \, 
\left(
\begin{array}{cc}
\epsilon_{0} - \mu + t T_{e}(\gmathbf{k}) + U(m_{1}-m_{2}) & - itT_{o}(\gmathbf{k})      \\
i t T_{o}(\gmathbf{k}) & \epsilon_{0} - \mu + t T_{e}(\gmathbf{k}) + U(m_{1}+m_{2})      \\
\end{array}
\right) \,\, .                                                          \label{Hamiltonian2}
\end{equation}

\noindent
In the above expressions, $\epsilon_{0}$ is the site energy, $\mu$ the chemical
potential, $m_{1}$ and $m_{2}$ represent the occupation number and magnetization 
along the local quantization axis, respectively, and $T_{e/o}(\gmathbf{k})$ read 
\cite{AS-91}

\begin{equation}
T_{e}(\gmathbf{k}) \, = \, 2 \left[ \cos k_{x} \, \cos \left(\frac{Q_{x}}{2}\right) 
+ \cos k_{y} \, \cos \left(\frac{Q_{y}}{2}\right) \right]                        \label{T-e}
\end{equation}

\begin{equation}
T_{o}(\gmathbf{k}) \, = \, 2 \left[ \sin k_{x} \, \sin \left(\frac{Q_{x}}{2}\right) 
+ \sin k_{y} \, \sin \left(\frac{Q_{y}}{2}\right) \right]  \,\, .                \label{T-o}
\end{equation}
                 
\noindent
The eigenvalues and eigenvectors of the matrix (\ref{Hamiltonian2}) are thus given by: 
                
\begin{equation}
\epsilon_{r}(\gmathbf{k}) \, = \, \epsilon_{0} - \mu + t T_{e}(\gmathbf{k}) + U m_{1}
+ (-1)^{r} \sqrt{ U^{2}m^{2}_{2} + t^{2}T_{o}(\gmathbf{k})^{2} }
                                                                         \label{eigenvalues}
\end{equation}

\noindent
($r=1,2$) and

\begin{equation}
W_{1}(\gmathbf{k}) =
\frac{1}{N_{1}(\gmathbf{k})}
\left(
\begin{array}{c}
1  \\
- i \frac{\frac{t}{U} \frac{1}{m_{2}} T_{o}(\gmathbf{k})}
{ 1 + \sqrt{1+\left(\frac{t}{U}\right)^{2}\frac{1}{m^{2}_{2}} T_{o}(\gmathbf{k})^{2}}}
\end{array}
\right)                                                                 \label{eigenvector1}
\end{equation}

 \begin{equation}
W_{2}(\gmathbf{k}) =
\frac{1}{N_{2}(\gmathbf{k})}
\left(
\begin{array}{c}
- i \frac{\frac{t}{U} \frac{1}{m_{2}} T_{o}(\gmathbf{k})}
{ 1 + \sqrt{1+\left(\frac{t}{U}\right)^{2}\frac{1}{m^{2}_{2}} T_{o}(\gmathbf{k})^{2}}} \\
1
\end{array}
\right)                                                                 \label{eigenvector2}
\end{equation}

\noindent
where $N_{r}(\gmathbf{k})$  stands for the normalization factor.

The parameters $m_{1}$, $m_{2}$, and $\gmathbf{Q}$ of the mean-field Hamiltonian
(\ref{Hamiltonian1}) are obtained, as usual, by minimizing the average value of the 
Hamiltonian with respect to the parameters themselves.
One obtains:

\begin{equation}
m_{1} = \frac{1}{2\gmathcal{N}} \sum_{\gmathbf{k}}^{BZ} \sum_{r} \,
f_{F}(\epsilon_{r}(\gmathbf{k})) = \frac{1 + \delta}{2}  \,\, ,                    \label{m1}
\end{equation}

\noindent
where $f_{F}(\epsilon)$ is the (zero-temperature) Fermi function and $\delta$
is the \emph{doping parameter\/},
            
\begin{equation}
m_2 = \frac{1}{2\gmathcal{N}} \sum_{\gmathbf{k}}^{BZ} \sum_{\xi} \sum_{r} \, \xi \,
           W^{\dagger}_{r,\xi}(\gmathbf{k}) \, W_{\xi,r}(\gmathbf{k}) \,
       f_{F}(\epsilon_{r}(\gmathbf{k}))  \, ,                                      \label{m2}
\end{equation} 

\noindent
and

\begin{equation}
\sum_{\gmathbf{k}}^{BZ} \sum_{\xi,\xi'} \,
\vec{\nabla}_{\gmathbf{Q}} \gmathcal{H}_{\xi,\xi'}(\gmathbf{k}) \, 
\sum_{r} W^{\dagger}_{r,\xi}(\gmathbf{k}) \,  W_{\xi',r}(\gmathbf{k}) \,
 f_{F}(\epsilon_{r}(\gmathbf{k})) \, = \, 0 \,\, .                       \label{consistency3}
\end{equation}
             
In the following, we shall restrict to the \emph{diagonal solution\/} 
$\gmathbf{Q}=Q(1,1)$, since it is known to be favored \emph{for sufficiently 
small values\/} of $t/U$.\cite{AS-91}
Accordingly, at the order we are considering of the small parameter $t/U$ 
we expand formally:

\begin{eqnarray}
\epsilon_{r}(\gmathbf{k}) & = & U 
\left[ \epsilon_{r}^{(0)}(\gmathbf{k})-\mu^{(0)}+\left(\frac{t}{U}\right) \right.
             (\epsilon_{r}^{(1)}(\gmathbf{k})-\mu^{(1)})           \nonumber  \\
& + & \left. \left(\frac{t}{U}\right)^{2}(\epsilon_{r}^{(2)}(\gmathbf{k})-\mu^{(2)}) + 
\cdots \right]                                                             \label{epsilon-2}
\end{eqnarray} 

\noindent
as well as
              
\begin{equation}
m_{2} = m_{2}^{(0)} + \left(\frac{t}{U}\right) m_{2}^{(1)}
      + \left(\frac{t}{U}\right)^{2} m_{2}^{(2)}
      + \cdots  \, ,                                                            \label{m2-2}
\end{equation}

\noindent
while, at the relevant order we can take

\begin{equation}
W_{1}(\gmathbf{k}) = \frac{1}{N_{r}(\gmathbf{k})} \left(
\begin{array}{c}
                 1                                                     \\
                -i \frac{t}{U} \frac{T_{o}(\gmathbf{k})}{2 m_{2}^{(0)}} \\
\end{array} 
\right)                                                                           \label{W1}
\end{equation}

\begin{equation}
W_{2}(\gmathbf{k}) = \frac{1}{N_{r}(\gmathbf{k})} \left(
\begin{array}{c}
                 -i \frac{t}{U} \frac{T_{o}(\gmathbf{k})}{2 m_{2}^{(0)}} \\
                  1                                                     \\
\end{array} 
\right)                                                                           \label{W2}
\end{equation}

\noindent
where

 \begin{equation}
\frac{1}{N_{r}(\gmathbf{k})^{2}}  = 1 - \left(\frac{t}{U}\right)^{2} 
\left(\frac{T_{o}(\gmathbf{k})}{2 m_{2}^{(0)}}\right)^{2} + 
\gmathcal{O}\left(\left(\frac{t}{U}\right)^{3}\right)                   \label{normalization}
\end{equation}

\noindent
is independent from $r$.
Note that it is sufficient to retain the lowest-order term $m_{2}^{(0)}$ in the above 
equations.
The parameter $m_{1}$, on the other hand, is given by Eq.~(\ref{m1}) and is 
thus formally independent from $t/U$ (we anticipate, however, that the doping 
parameter $\delta$ will turn out to be at most of the order $t/U$ for our 
expansions to be internally consistent).

The coefficients $\epsilon_{r}^{(n)}(\gmathbf{k})$ of Eq.~(\ref{epsilon-2}) with 
$n=0,1,2,\cdots$ can be readily obtained from the expression (\ref{eigenvalues}) 
(where we may set $\epsilon_{0}=0$ for simplicity) in terms of the $
m_{2}^{(n)}$ (for given $\gmathbf{Q}$).
The value of $\mu^{(0)}$ can also be readily obtained in terms of $\epsilon_{r}^{(0)}$
(cf. Appendix ~B).
The remaining coefficients of the expansions (\ref{epsilon-2}) and (\ref{m2-2})
can further be determined by solving the coupled equations for the self-consistency
parameters according to the methods developed in the Appendices B and C.
In particular, for $\delta>0$ we obtain: 

\begin{eqnarray}                               
m_{2}^{(0)} & = & \frac{1}{2}(1-\delta)                      \nonumber  \\
m_{2}^{(1)} & = & 0                                          \nonumber  \\
m_{2}^{(2)} & = & - 4 \sin^{2}(Q/2) + \gmathcal{O}(\delta) \,\, ,             \label{expan-1}
\end{eqnarray}

\begin{eqnarray}
\mu^{(0)} - m_{2}^{(0)} & = & \frac{1}{2}(1+\delta)            \nonumber \\
\mu^{(1)} - m_{2}^{(1)} & = &
\cos(Q/2)[4 - 4 \pi \delta + \gmathcal{O}(\delta^{2})]          \nonumber \\
\mu^{(2)} - m_{2}^{(2)} & = & 0 + \gmathcal{O}(\delta)  \,\, ,                \label{expan-2}
\end{eqnarray}

\noindent
and

\begin{eqnarray}
\epsilon_{r}^{(0)}(\gmathbf{k}) & = & \frac{1}{2} \left[(1+\delta) 
              + (-1)^{r}(1-\delta)\right]                       \nonumber  \\
\epsilon_{r}^{(1)}(\gmathbf{k}) & = & T_{e}(\gmathbf{k})          \nonumber  \\
\epsilon_{r}^{(2)}(\gmathbf{k}) & = & (-1)^{r} \left[T_{o}^{2}(\gmathbf{k})
- 4 \, \sin ^{2}(Q/2)\right] + \gmathcal{O}(\delta) \,\, ,                    \label{expan-3}
\end{eqnarray}

\noindent
where for the spiral configuration we are considering $Q$ is determined by 
the condition

\begin{equation} 
\cos (Q/2) = \frac{- U \delta}{2t} + \gmathcal{O}\left(t/U\right)                 \label{cos} 
\end{equation}

\noindent
with $\delta \leq 2(t/U) + \gmathcal{O}((t/U)^{2})$, as anticipated.
(The other allowed solution $\sin (Q/2) = 0$ describes instead the ferromagnetic case.)
As discussed in Appendix B, the above results have been obtained with the further
assumption that $\delta$ is small enough but not infinitesimal, i.e., $\delta$
satisfies the condition $(t/U)^{2}<<\delta$.
The case $\delta=0$, on the other hand, can be considered separately. 
Note that Eq.~(\ref{cos}) implies that in the spiral phase $\delta$ is at most of the order 
$t/U$. 
This property has to be taken into  account to get a consistent expansion up to the desired
order in $t/U$.

\vspace{0.8cm}
\begin{center}
\begin{large}
{\bf B. Susceptibilities and spin-wave dispersion}
\end{large}
\end{center}
\vspace{0.8cm}

Before performing the $t/U$ expansion of the matrix elements of the non-interacting 
susceptibility tensor (\ref{da}) to get the spin-wave dispersion relation, it is convenient 
to exploit some symmetry properties that reduce the number of matrix elements to
be considered.
Specifically, from the property
 
\begin{equation}
\gmathcal{F}_{r,r'}(\gmathbf{k},\gmathbf{k}-\gmathbf{q},\omega) = 
\gmathcal{F}_{r',r}(\gmathbf{k}-\gmathbf{q},\gmathbf{k},-\omega)              \label{symmetry-F}
\end{equation}

\noindent
it follows that:

\begin{equation}
\gmathcal{X}_{0}^{\alpha,\beta}(\gmathbf{q},\omega) = 
\gmathcal{X}_{0}^{\beta,\alpha}(-\gmathbf{q},-\omega)  \,\, .               \label{symmetry-X}
\end{equation}

\noindent
By direct inspection it can also be verified that:  

\begin{eqnarray}
\gmathcal{X}_{0}^{0,-}(\gmathbf{q},\omega)     & = &
- \gmathcal{X}_{0}^{+,0}(\gmathbf{q},\omega)   \, ,             \nonumber \\      
\gmathcal{X}_{0}^{z,-}(\gmathbf{q},\omega)     & = &
- \gmathcal{X}_{0}^{+,z}(\gmathbf{q},\omega)   \, ,             \nonumber \\      
\gmathcal{X}_{0}^{+,+}(\gmathbf{q},\omega)     & = &
\gmathcal{X}_{0}^{-,-}(\gmathbf{q},\omega)     \, .                           \label{relate-X} 
\end{eqnarray}

\noindent
In this way, the matrix (\ref{definition-M}) acquires the simplified form:

\begin{equation}
M(\gmathbf{q},\omega) =
\left(
\begin{array}{cccc}
1-a(\gmathbf{q},\omega)&ib(\gmathbf{q},\omega)&-ib(-\gmathbf{q},-\omega)&c(\gmathbf{q},\omega)\\
ib(\gmathbf{q},\omega)&1+d(\gmathbf{q},\omega)&e(\gmathbf{q},\omega)&if(\gmathbf{q},\omega)\\
-ib(-\gmathbf{q},-\omega)&e(\gmathbf{q},\omega)&1+d(-\gmathbf{q},-\omega)&-if(-\gmathbf{q},-\omega)\\
-c(\gmathbf{q},\omega)&-if(\gmathbf{q},\omega)&if(-\gmathbf{q},-\omega)&1+g(\gmathbf{q},\omega)\\
\end{array}
\right)                                                          \label{matrix-M-simplified}
\end{equation}

\noindent
where we have set

\begin{eqnarray}
a(\gmathbf{q},\omega) & = & 2 \, U \, \gmathcal{X}_{0}^{0,0}(\gmathbf{q},\omega)          \nonumber \\
b(\gmathbf{q},\omega) & = & -2 \, i \, U \, \gmathcal{X}_{0}^{0,-}(\gmathbf{q},\omega)    \nonumber \\
c(\gmathbf{q},\omega) & = & 2 \, U \, \gmathcal{X}_{0}^{0,z}(\gmathbf{q},\omega)          \nonumber \\
d(\gmathbf{q},\omega) & = & 2 \, U \, \gmathcal{X}_{0}^{+,-}(\gmathbf{q},\omega)          \label{a-g} \\
e(\gmathbf{q},\omega) & = & 2 \, U \, \gmathcal{X}_{0}^{+,+}(\gmathbf{q},\omega)          \nonumber \\
f(\gmathbf{q},\omega) & = & -2 \, i \, U \, \gmathcal{X}_{0}^{+,z}(\gmathbf{q},\omega)    \nonumber \\
g(\gmathbf{q},\omega) & = & 2 \, U \, \gmathcal{X}_{0}^{z,z}(\gmathbf{q},\omega)          \nonumber \,\, .
\end{eqnarray}

Entering then the expansions (\ref{W1})-(\ref{normalization}) for the eigenvectors of 
the mean-field Hamiltonian (with $m_{2}^{(0)}$ given by Eq.~(\ref{expan-1})) into the
definition (\ref{du}), we obtain the expressions for the relevant matrix elements of 
the non-interacting susceptibility tensor (\ref{da}) reported in Appendix B at the order 
in $t/U$ we are considering.
Utilizing further the method developed in Appendix C to perform the $\gmathbf{k}$ summation 
when the doping parameter $\delta$ is small, we obtain eventually the following expressions 
for the matrix elements (\ref{a-g}):

\begin{eqnarray}
a(\gmathbf{q},\omega) & = & a(\gmathbf{q}) = \frac{-1}{2 \,\cos (Q/2) \, (\cos q_{x}+\cos q_{y}-2)}
\, \frac{\delta U}{t} + \gmathcal{O}\left(t/U\right) \, ,      \nonumber \\
b(\gmathbf{q},\omega) & = & b(\gmathbf{q}) = \frac{- \sin (Q/2) \, (\sin q_{x}+\sin q_{y})}
{\sqrt{2} \, \cos (Q/2) \, (\cos q_{x}+\cos q_{y}-2)} \, \delta 
+ \gmathcal{O}\left(\left(t/U\right)^{2}\right) \, ,                            \nonumber \\
c(\gmathbf{q},\omega) & = & c(\gmathbf{q}) = - a(\gmathbf{q},\omega) 
+ \gmathcal{O}\left(t/U\right) \, ,                            \nonumber \\
e(\gmathbf{q},\omega) & = & e(\gmathbf{q}) = - 4 \,\left(\frac{t}{U}\right)^{2} 
\, \sin ^{2}(Q/2) \, (\cos q_{x}+\cos q_{y}) +
\gmathcal{O}\left(\left(t/U\right)^{3}\right) \, ,                              \nonumber \\
f(\gmathbf{q},\omega) & = & f(\gmathbf{q}) = b(\gmathbf{q},\omega) 
+ \gmathcal{O}\left(\left(t/U\right)^{2}\right) \, ,                            \nonumber \\
g(\gmathbf{q},\omega) & = & g(\gmathbf{q}) = a(\gmathbf{q},\omega) 
+ \gmathcal{O}\left(t/U\right) \, ,                                         \label{a-b-c-f-g}
\end{eqnarray}

\noindent
as well as

\begin{eqnarray}
d(\gmathbf{q},\omega) & \equiv &  
-1 - \tilde{\omega} + \alpha(\gmathbf{q}) + 
\gmathcal{O}\left(\left(t/U\right)^{3}\right)                      \nonumber \\               
& = & - 1 - \tilde{\omega} + \left(\frac{t}{U}\right)^{2} \, 4 \, 
\cos ^{2}(Q/2) \, (\cos q_{x}+\cos q_{y}-2) + 
8 \left(\frac{t}{U}\right) ^{2} \sin ^{2}(Q/2)                    \nonumber \\  
& + &  2 \, \left(\frac{t}{U}\right) \, \delta \, \cos (Q/2) \,
(\cos q_{x}+\cos q_{y}-2)                                         \nonumber \\     
& - &   2 \, \left(\frac{t}{U}\right) \, \delta \,
\frac{\sin ^{2}(Q/2) \, (\sin q_{x}+\sin q_{y})^{2}}
     {\cos (Q/2) \, (\cos q_{x}+\cos q_{y}-2)} \,   
+ \, \gmathcal{O}\left(\left(t/U\right)^{3}\right)                                \label{d-1}
\end{eqnarray}

\noindent
where $\tilde{\omega}\equiv\omega/U$ will turn out to be of order $t^2/U^2$ at the spin-wave poles.

To obtain the above expressions, we have considered only the real part of the
functions (\ref{A-bubble-definition}). 
This is definitely possible for every pairs of bands when we restrict to values
of $\gmathbf{q}$ such that $|\gmathbf{q}| \gg k_{F}$, where $k_{F}$ (by our 
definition) is the maximum value of the function $k(\phi)$ introduced
in Appendix C, which coincides with the Fermi momentum  for small $\delta$.
Since we have shown in the same Appendix that $k(\phi) \sim \sqrt{\delta} \sim
\sqrt{t/U}$, taking the $\gmathbf{q} \rightarrow 0$ limit implies letting
$t/U$ to vanish before $\gmathbf{q}$.

Note also that, although the expressions (\ref{a-b-c-f-g}) and (\ref{d-1}) have been 
calculated at different orders in $t/U$, the ensuing expression for the determinant 
of the matrix (\ref{matrix-M-simplified}) is obtained consistently at the fourth order
in $t/U$, as required for the frequency of the spin-wave mode to be of the order
of $t^{2}/U$.
In fact, by writing the determinant explicitly we obtain:

\begin{eqnarray}
& & \left[ \left(1 - a(\gmathbf{q})\right) \left(1 + a(\gmathbf{q})\right) + a^{2}(\gmathbf{q}) 
+ \gmathcal{O}\left(t/U\right) \right]                                                \nonumber \\    
& & \times  \left[ \left(- \tilde{\omega}+\alpha(\gmathbf{q})\right)
                     \left(\tilde{\omega}+\alpha(\gmathbf{q})\right)
- e^{2}(\gmathbf{q}) + \gmathcal{O}\left(\left(t/U\right)^{5}\right) \right]           \nonumber \\     
& & + 2 \, \left[1-a(\gmathbf{q}) + \gmathcal{O}\left(t/U\right) \right]      
\left[ b^{2}(\gmathbf{q}) + \gmathcal{O}\left(\left(t/U\right)^{3}\right) \right]      
\left[ e(\gmathbf{q})-\alpha(\gmathbf{q}) 
                   + \gmathcal{O}\left(\left(t/U\right)^{3}\right) \right]            \nonumber \\     
& & -2 \, \left[1+a(\gmathbf{q}) + \gmathcal{O}\left(t/U\right)\right]      
\left[ b^{2}(\gmathbf{q}) + \gmathcal{O}\left(\left(t/U\right)^{3}\right) \right]      
\left[ e(\gmathbf{q})-\alpha(\gmathbf{q}) 
                   + \gmathcal{O}\left(\left(t/U\right)^{3}\right) \right]            \nonumber \\     
& & + \, 4 \left[ a(\gmathbf{q}) + \gmathcal{O}\left(t/U\right) \right]      
\left[ b^{2}(\gmathbf{q}) + \gmathcal{O}\left(\left(t/U\right)^{3}\right) \right]      
\left[ e(\gmathbf{q})-\alpha(\gmathbf{q}) 
               + \gmathcal{O}\left(\left(t/U\right)^{3}\right)\right] \, = \, 0  \, . \nonumber \\     
                                                                 \label{original-dispersion}
\end{eqnarray} 

\noindent
Note that the last three terms on the left-hand side add up to zero at the fourth order
in $t/U$ we are considering.
We are thus left with the expression

\begin{equation}
\tilde{\omega}^{2} - \alpha^{2}(\gmathbf{q}) + e^{2}(\gmathbf{q}) 
+ \gmathcal{O}\left(\left(t/U\right)^{5}\right) \, = \, 0 \, ,                \label{omega-1}
\end{equation}

\noindent
yielding

\begin{equation}
\omega^{2}(\gmathbf{q}) = U^{2} \left( \alpha^{2}(\gmathbf{q}) - e^{2}(\gmathbf{q}) \right) 
+ \gmathcal{O}\left(t^{5}/U^{3}\right)  \,\, .                                \label{omega-2}
\end{equation}

\noindent
Taking into account the expressions for $\alpha(\gmathbf{q})$ and $e(\gmathbf{q})$ given
above, we get eventually the desired spin-wave dispersion relation, in the form:

\begin{eqnarray}
\omega^{2}(\gmathbf{q}) & = & 16 \, \frac{t^{4}}{U^{2}} \,\cos Q \, (\cos q_{x}+\cos q_{y}-2) 
\, \left[\cos q_{x}+\cos q_{y}-2 \cos Q \right]                             \nonumber \\
& + &  16 \, \frac{t^{3}}{U} \, \delta \, \left[\cos (Q/2) \, (\cos q_{x}+\cos q_{y}-2) -
\frac{\sin ^{2}(Q/2) \, (\sin q_{x}+\sin q_{y})^{2}}{\cos (Q/2) \, (\cos q_{x}+\cos q_{y}-2)}     
\right]                                                                     \nonumber \\
& \times & \left[\cos ^{2}(Q/2) \, (\cos q_{x}+\cos q_{y}-2) + 2 \, \sin ^{2}(Q/2)
\right]                                                                     \nonumber \\
& + &   4 \, \delta^{2} \, t^{2} \, \left[\cos (Q/2) \, (\cos q_{x}+\cos q_{y}-2) -
\frac{\sin ^{2}(Q/2) \, (\sin q_{x}+\sin q_{y})^{2}}{\cos (Q/2) \, (\cos q_{x}+\cos q_{y}-2)}
\right]^{2}  \,\, .                                                          \nonumber \\
                                                                    \label{final-dispersion}
\end{eqnarray}                                                                          

\noindent
To obtain the physical dispersion relation, however, there remains to enter in the
above expression the relation among $Q$, $\delta$, and $t/U$ as given by the mean-field
condition (\ref{cos}), as discussed in the next Section.


\vspace{2cm}
\section{Results and discussion}

In this Section, we discuss the physical consequences of the dispersion
relation (\ref{final-dispersion}) for spin-wave excitations over an 
incommensurate (diagonal) spiral magnetic configuration of a 
two-dimensional Hubbard Hamiltonian.

We have already remarked that the dispersion relation (\ref{final-dispersion})  
is not yet in its final form, since the connection among $Q$, $\delta$, 
and $t/U$ needs still to be specified.
Before considering the general case, however, it is interesting to recover from
Eq.~(\ref{final-dispersion}) the spin-wave dispersion relations corresponding
to the limiting cases of an antiferromagnet and of a ferromagnet.

When $\delta =0$, Eq.~(\ref{B-zero-2}) yields $\cos (Q/2)=0$, that is 
$\gmathbf{Q}= \gmathbf{Q}_{AF}=(\pi,\pi)$. 
In this case we obtain from Eq.~(\ref{final-dispersion}) (by setting $\delta=0$ 
identically therein):

\begin{equation} 
\omega(\gmathbf{q}) = J_{\gmathrm{eff}}^{(AF)} \,       
      \left[ 4- \left(\cos q_{x}+\cos q_{y}\right)^{2} \right]^{\frac{1}{2}}
                                                                          \label{disp-rel-2} 
\end{equation}

\noindent
with $J_{\gmathrm{eff}}^{(AF)} = 4 t^{2}/U$.
This result coincides with the spin-wave dispersion relation of a two-dimensional 
Heisenberg \emph{antiferromagnet\/} at leading order in $t/U$.\cite{AF} 

When  $\gmathbf{Q}=(2\pi,2\pi)$ and $\delta$ arbitrary, we obtain instead
from Eq.~(\ref{final-dispersion})

\begin{equation}
\omega(\gmathbf{q}) = \frac{4 t^{2}}{U} \left|\left( 1 - \frac{\delta U}{2 t} \right)\right|
                  (2-\cos q_{x}-\cos q_{y})  \, ,                         \label{disp-rel-3}
\end{equation}
                               
\noindent                               
which coincides with the spin-wave dispersion relation of a nearest-neighbor 
Heisenberg \emph{ferromagnet\/} with 
$J_{\gmathrm{eff}}^{(F)} = 4 t^{2}/U \left[1-\delta U/(2t)\right]$.
Note that $J_{\gmathrm{eff}}^{(F)}$ is negative since the (mean-field) ferromagnetic 
solution is actually stable when $\delta \geq 2t/U$.

In the general case of a (diagonal) spiral spin configuration, the relation
among $Q$, $\delta$, and $t/U$ is given by Eq.~(\ref{B-zero-2}), that is,
$\delta = - (2t/U) \cos Q/2 + \gmathcal{O}((t/U)^{2})$ at the leading order in
$t/U$. 
Eliminating $Q$ in favor of $\delta$ and $t/U$ via this relation in 
Eq.~(\ref{final-dispersion}), we obtain eventually the following expression:

\begin{equation}
\omega(\gmathbf{q}) = J_{\gmathrm{eff}} \, 
\left\{ \left[ 2-\frac{(\sin q_{x}+\sin q_{y})^{2}}{2-\cos q_{x}-\cos q_{y}} \right]^{2} \right.
\left. - \, (\cos q_{x}+\cos q_{y})^{2} \phantom{\frac{L}{R}} \right\}^{\frac{1}{2}}       
                                                                          \label{disp-rel-4}
\end{equation}

\noindent
where now the effective exchange integral is given by

\begin{equation} 
J_{\gmathrm{eff}} = \frac{4 t^{2}}{U} \, \sin ^{2}(Q/2) = \frac{4 t^{2}}{U}
\left(1-\left(\frac{U}{2t}\delta\right)^{2}\right) 
= \frac{4 t^{2}}{U} \left(1-\left(\frac{\delta}{\delta_{c}}\right)^{2}\right)   
                                                                             \label{delta-c}    
\end{equation}           

\noindent
with $\delta_{c} \equiv 2t/U$.
Equation (\ref{disp-rel-4}) constitutes the main result of this paper.
Note that when $\delta$ reaches the \emph{critical\/} value $\delta_{c}$,
$J_{\gmathrm{eff}}$ vanishes.
Past this value, the spiral solution evolves into the ferromagnetic solution, 
which becomes the stable solution (at the mean-field level).

The real and imaginary parts of $\omega(\gmathbf{q})$ (in units of $J_{\gmathrm{eff}}$),
as obtained from the analytic expression (\ref{disp-rel-4}), are plotted in 
Figs.~1(a) and 1(b), respectively, over the two-dimensional Brillouin
zone.
\begin{figure}[htb] 
\par \vspace*{.5cm} \par
   \centerline{
\psfig{file=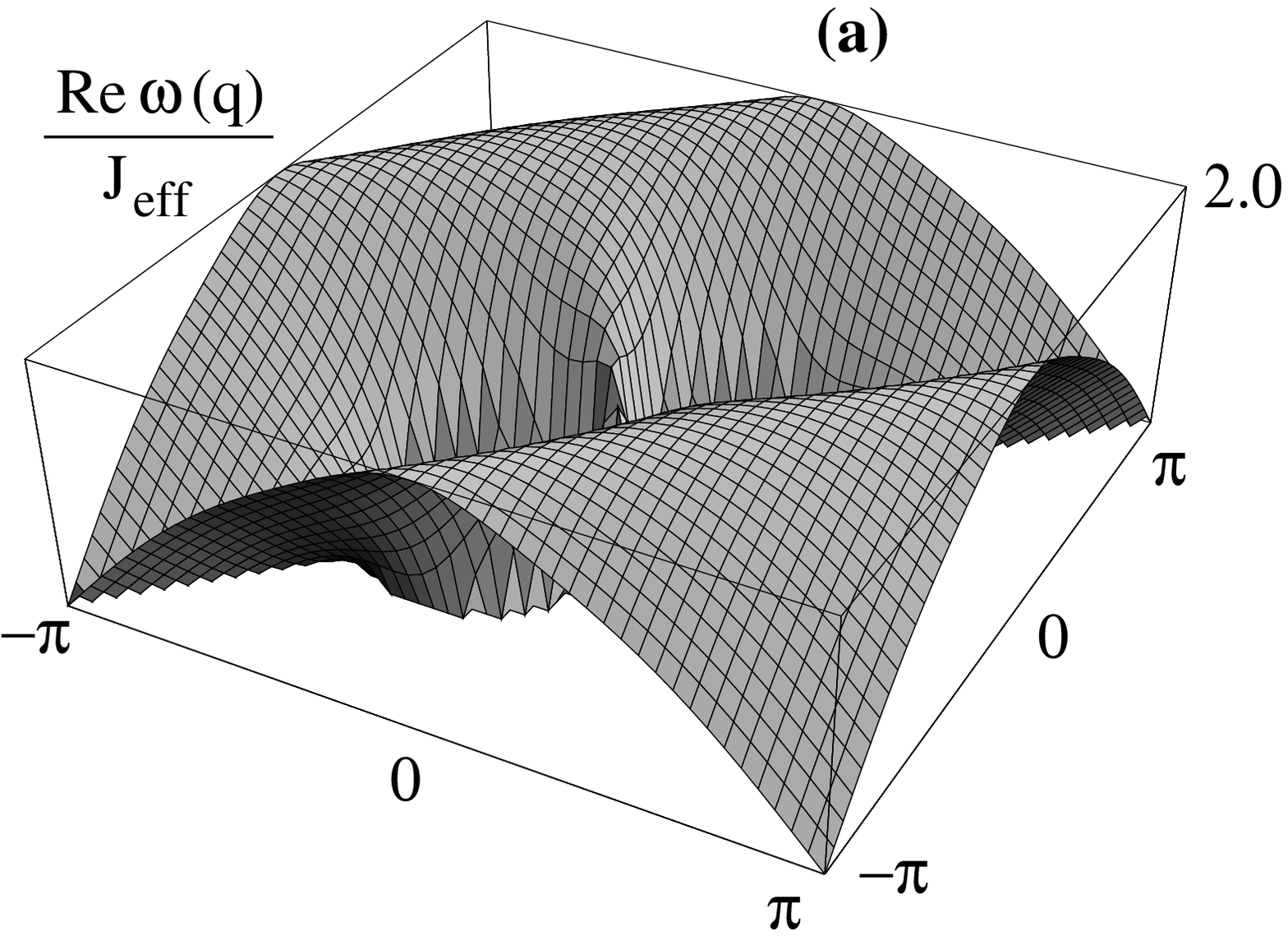,width=8.cm}
\psfig{file=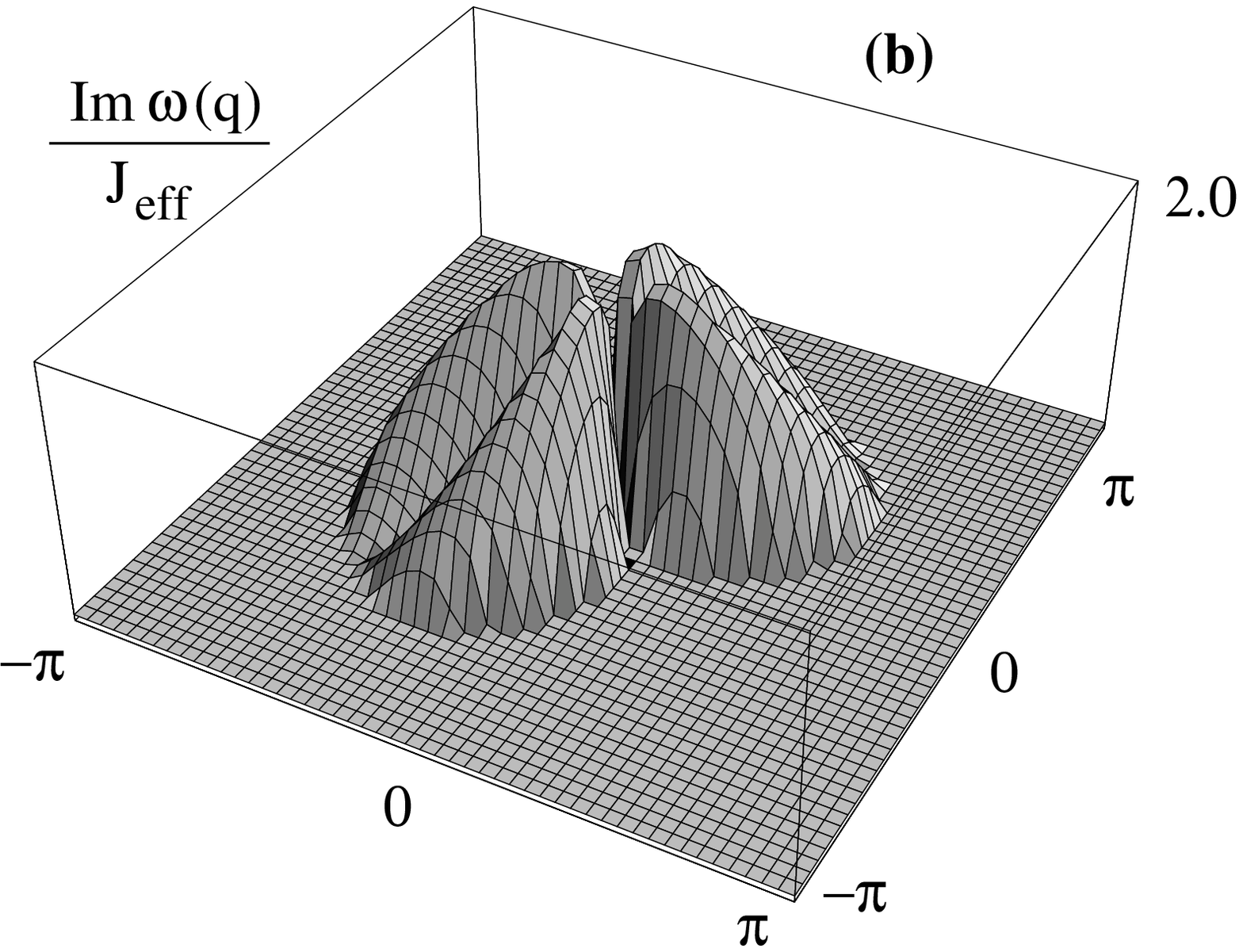,width=8.cm}
}
\caption{
 Real (a) and imaginary (b) part of the spin-wave dispersion
relation (\protect{\ref{disp-rel-4}}) 
(in units of $J_{\gmathrm{eff}}$), obtained with a 
diagonal-spiral spin configuration background.
Note that the $Q$-dependence is contained in $J_{\gmathrm{eff}}$ only.
}
\par \vspace*{.5cm} \par
 \end{figure}

It is clear from the expression (\ref{disp-rel-4}) that $\omega(\gmathbf{q})$ is
\emph{either\/} real \emph{or\/} purely imaginary, so that, when its real part is 
nonvanishing, its imaginary part vanishes identically, and viceversa.
Actually, this is true  in the region of the Brillouin zone where our expansion holds,
namely, for $|\gmathbf{q}|\gg k_{F}$  [cf. the discussion below Eq.~(\ref{d-1})].
For $|\gmathbf{q}|<k_{F}$, the spin-wave dispersion acquires a
damping due to the mixing with the particle-hole continuum.\cite{footnote-1}
An exception is represented by the line $q_{x}=q_{y}$, along which the real
and imaginary parts vanish simultaneously.
The softening of the dispersion relation along the whole line $q_{x}=q_{y}$ and not
only when $\gmathbf{q}=0$ and $\gmathbf{q}=\gmathbf{Q}$ (as one would expect on general
grounds in the presence of a spiral spin configuration \cite{Brenig}) corresponds to 
the fact that $\omega(\gmathbf{q})$ given by Eq.~(\ref{disp-rel-4}) can be cast in the 
form $\omega(\gmathbf{q}) = f(\gmathbf{Q}) \, g_{\hat{Q}}(\gmathbf{q})$;
as $\omega(\gmathbf{q}=\gmathbf{Q}) = 0$ invariably, it follows that 
$g_{\hat{Q}}(\gmathbf{q}=\gmathbf{Q}) = 0$ since $g_{\hat{Q}}(\gmathbf{q})$ does not 
depend on $|\gmathbf{Q}|$.
This implies that $g_{\hat{Q}}(q_{x}=q_{y}) \equiv 0$ identically.

One expects this result to be modified, however, at higher order in $t/U$.
In this respect, it is interesting to compare with the results obtained by
Brenig \cite{Brenig} by solving numerically the condition (\ref{zero}) directly,
without performing the expansion in the small parameter $t/U$.
One sees, in particular, from Fig.~6 of Brenig's paper (obtained for $\delta = 0.075$)
that $\omega(\gmathbf{q})$ remains indeed finite along $q_{x}=q_{y}$ already when $t/U = 0.1$.

\begin{figure}[htb] 
\par \vspace*{.5cm} \par
   \centerline{
\psfig{file=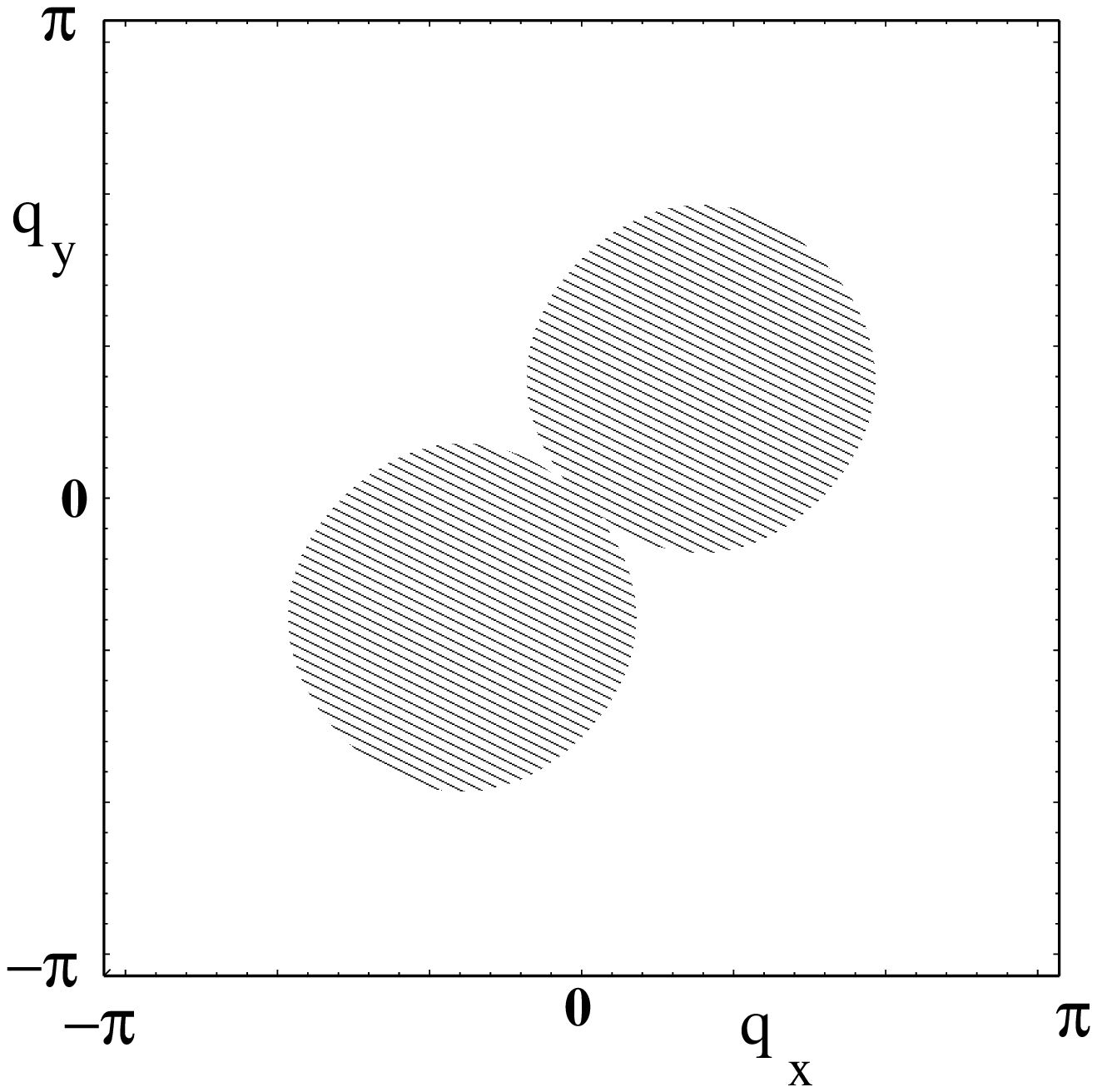,width=6.cm}
}
\caption{
Region of the two-dimensional Brillouin zone where the
spin-wave dispersion is overdamped (shaded area).
Note that this region does not depend on $Q$, and that the region about 
the center of the Brillouin zone has been excluded by construction.
}
\par \vspace*{.5cm} \par
 \end{figure}

Figure 2 shows the region of the Brillouin zone (shaded area) where $\omega(\gmathbf{q})$ 
is \emph{overdamped\/} (purely imaginary) (with the region about $\gmathbf{q}=0$ excluded
according to the argument given in Section 3).
A finite region of the Brillouin zone where the dispersion relation is overdamped is
also reported in Ref.~\cite{Brenig}, even though a direct comparison with our results
is not possible owing to the different ranges of the parameter $t/U$ explored.
This overdamping signals an instability of the system (due to the merging
into the particle-hole continuum) toward a different ground state, reflecting possibly
a more complicated long-range spin (and charge) structure than the spin-spiral one 
considered in the present paper. 
Nonetheless, we expect on physical grounds that close to the boundary of the Brillouin 
zone (where overdamping of spin waves does not occur in our calculation) the spin-wave 
spectrum with small wavelength obtained by our approach would survive the inclusion of 
more complicated long-range spin structures.

It is also interesting to compare the form of the dispersion relation (\ref{disp-rel-4}) 
with the spin-wave dispersion relation obtained with the Heisenberg model including second 
and third nearest-neighbor couplings, for the same value of the incommensurability wave 
vector $Q$ (cf. Appendix D).
This comparison is shown in Fig.~3 along the symmetry lines of the Brillouin zone and 
evidences marked differences between the two dispersion relations.

\begin{figure}[htb] 
\par \vspace*{.5cm} \par
   \centerline{
\psfig{file=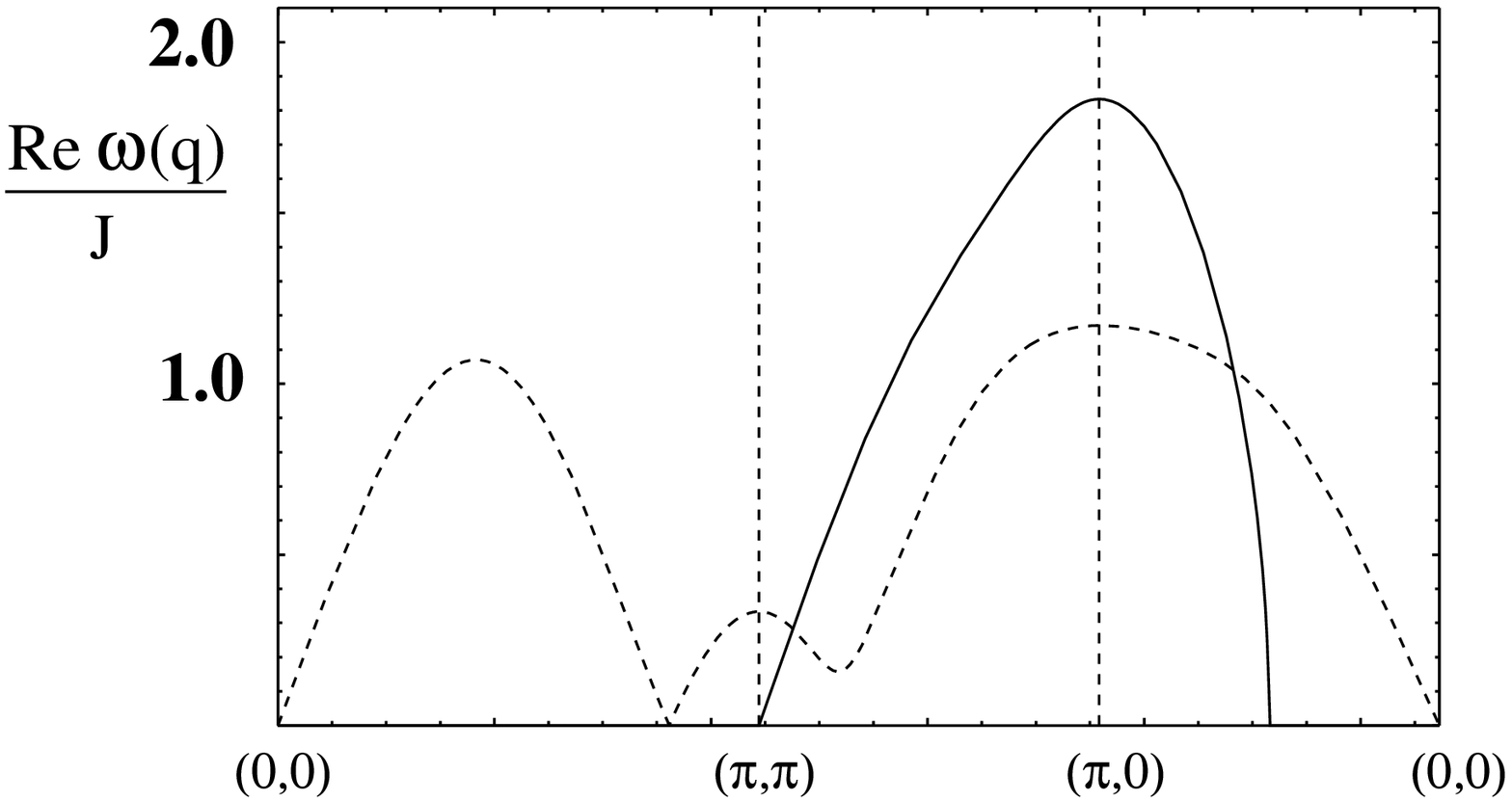,width=10.cm}
}
\caption{
Comparison of the spin-wave dispersion relations along the
symmetry directions of the Brillouin zone, as obtained for the itinerant model 
(full line) and for the Heisenberg model with second ($J_{2}/J_{1}=0.2$) and 
third ($J_{3}/J_{1}=0.2$) nearest-neighbor couplings (broken line). 
Both dispersions correspond to the same value of $Q=2.556$ and are 
plotted in units of $J=4 t^2/U$ and $J_1$, respectively.
}
\par \vspace*{.5cm} \par
 \end{figure}

Returning to the dispersion relation (\ref{disp-rel-4}) for $\delta \ne 0$, we emphasize 
that its functional form could not be obtained from the dispersion relation (\ref{disp-rel-2}) 
valid when $\delta = 0$, by simply modifying the numerical values of the exchange integral 
therein. 
Nor, it would be sufficient to include a \emph{finite number\/} of exchange integrals 
in the Heisenberg model to account for the finite doping, owing to 
the presence of an RKKY-type term in Eq.~(\ref{disp-rel-4}) which contains trigonometric 
functions in the denominator rather than in the numerator only (cf. Appendix E).
Recall, in fact, that for practical purposes the Heisenberg model can be regarded as a 
\emph{fitting model\/} which could, in principle, reproduce \emph{any\/} type of spin-wave 
dispersion relation, provided a sufficiently large number of terms associated with 
distant neighbors were included.\cite{AM}
In addition, it appears fair to say that it would have been certainly difficult to guess 
\emph{a priori\/} the functional form of the dispersion relation (\ref{disp-rel-4}), by 
fitting the numerical results for the itinerant model onto the Heisenberg model extended
to a large number of neighbors.  

It is also interesting to comment on the effective exchange integral (\ref{delta-c})
retaining the $t^{2}/U$ dependence of the nearest-neighbor Heisenberg model, even when 
couplings between far apart neighbors are considered.
From a perturbative point of view, when $U>>t$ the magnetic interaction between a 
given lattice site and far apart neighbors is provided by the \emph{mobility\/}
of the holes in the doped configurations.
In this respect, one should consider all possible configurations with empty sites
distributed at random over the lattice sites, the mobility of the holes then resulting
by diagonalizing the Hamiltonian in this basis. 
As a consequence, the magnetic exchange coupling turns out to be proportional to 
$t^{2}/U$ at leading order, even for coupling between sites at arbitrary distances.

As anticipated in the Introduction, the form (\ref{disp-rel-4}) is somewhat hybrid
between the one obtained with a nearest-neighbor Heisenberg model (cf. Appendix D)
and the long-range RKKY interaction mediated by the conduction electrons (cf. Appendix E).
These two contributions to Eq.~(\ref{disp-rel-4}) cannot be separated in a
clear and unambiguous way.
However, it is possible to trace their origin by considering the expressions 
(\ref{plus-minus}) and (\ref{plus-plus}) for the relevant matrix elements 
of the susceptibility.

If one could set ``by hand'' $\gmathcal{F}_{22} = 0$ in those expressions, thus keeping 
only the interband terms, one would in fact obtain for the dispersion relation:

\begin{equation}
\omega (\gmathbf{q}) = J_{\gmathrm{eff}}  
\left[ 4 -(\cos q_{x}+\cos q_{y})^{2} \right]^{\frac{1}{2}}  \,\, ,              \label{HEI}
\end{equation}

\noindent
which corresponds to a nearest-neighbor Heisenberg antiferromagnet with exchange
coupling given by Eq.~(\ref{delta-c}).
Similarly, if one could set $\gmathcal{F}_{21}=\gmathcal{F}_{12}=0$ ``by hand'', thus 
keeping only the intraband terms, one would instead obtain an expression of the RKKY-type:

\begin{equation}
\omega(\gmathbf{q}) = J_{\gmathrm{eff}} \,\,\,
\frac{(\sin q_{x}+\sin q_{y})^{2}}{2-\cos q_{x}-\cos q_{y}}                     \label{RKKY}
\end{equation}

\noindent
with $J_{\gmathrm{eff}}$ still given by  Eq.~(\ref{delta-c}).
Note that in both cases the spectrum would be real and no overdamping would occur.
Our general result (\ref{disp-rel-4}) can then be cast in the following appealing 
form:

\begin{equation}
\omega(\gmathbf{q}) \, = \, \left\{ \left[ J(\gmathbf{q}=0) \, - \, 
\omega(\gmathbf{q})_{RKKY} \right]^{2} \, - \, J(\gmathbf{q})^{2} \right\}^{1/2}
                                                                              \label{hybrid}
\end{equation} 

\noindent
where $J(\gmathbf{q})=J_{\gmathrm{eff}} \, (\cos q_{x} + \cos q_{y})$ and  
$\omega(\gmathbf{q})_{RKKY}$ is given by the expression (\ref{RKKY}).
Note, in particular, that if one could set $\omega(\gmathbf{q})_{RKKY}=0$, the 
Heisenberg form (\ref{HEI}) would result from Eq.~(\ref{hybrid}).                                                            

A final comment on the possible comparison between the spin-wave spectrum we
have obtained and the available experimental data is in order.
That the classical spin-wave theory (possibly with quantum corrections \cite{co.ha.99}) 
can accurately describe the spin-wave spectrum 
over the whole Brillouin zone for the parent compounds of high-temperature
superconductors (zero doping) is well established at this point.\cite{spiral-data}
That when carriers are added (finite doping) incommensurate spin fluctuations
occur with rather short coherence length has also been well established in
several materials.\cite{spiral-data}
It is interesting to point out, in addition, that the observation of well-defined 
spin waves only close to the boundary of the Brillouin zone has been 
reported,\cite{Aeppli} consistently with what we have obtained by our calculation.
However, detailed comparison with our form (\ref{disp-rel-4}) of the spin-wave 
dispersion relation with the experimental data may not be possible, because our result 
is valid in the asymptotic limit  $U>> W$, where $W\sim 8 t$ is the bandwidth,
which may not be realized for real materials.


\vspace{2cm}
\section{Concluding remarks}

In this paper, we have studied the spin-wave spectrum for a two-dimensional
Hubbard Hamiltonian in the presence of an incommensurate spin-spiral phase
within the electronic RPA approximation in the broken-symmetry phase.
We have, in particular, obtained the analytic form of this spectrum at the
leading order in the small parameter $t/U$ of the Hubbard Hamiltonian.
Specifically, it has been possible to obtain the spectrum in a \emph{closed 
form\/} even in the presence of an \emph{incommensurate\/} structure, owing 
to the peculiar symmetry which is intrinsic to the spiral phase.
In this respect, starting from an alternative mean-field configuration
different from the spiral one (such as, for instance, a ``stripe'' structure) 
would have not enabled us to solve for the spin wave spectrum in a
\emph{closed form\/}.
By our approach, we have thus pushed the analytic results for the spin wave 
spectrum of an incommensurate structure as far as possible (apart, obviously, 
from including higher orders in the $t/U$ expansion which could still be 
done by our approach).

Alternatively, the spin-wave spectrum could have been obtained numerically
for any value of $t/U$ without performing the small $t/U$ expansion, as
already reported in Ref.~\cite{Brenig}.
In this way, however, it would have been rather difficult (if not impossible)
to arrive at the functional form (\ref{disp-rel-4}) for the spin-wave dispersion 
relation in the strong-coupling limit, which has an hybrid form between the
dispersion relation for a nearest-neighbor Heisenberg model and that obtained
within the (long-range) RKKY interaction in the presence of a finite doping.
Owing to the itinerant character of the system, it is, in fact, the presence 
of a band of metallic character (crossed by the Fermi level) which generates 
the RKKY magnetic interaction between the localized spins associated with the 
(filled) lower band.
This novel feature constitutes the main result of the present paper.
We have concluded accordingly that, even for small doping, the itinerant model 
we have considered results in a dispersion relation $\omega(\gmathbf{q})$ that 
cannot be effectively represented by the Heisenberg model, for which the couplings 
extend to a few neighbors only.
This occurs because the free carriers, associated with the itinerant character of
our starting Hamiltonian, necessarily introduce long-range RKKY-type magnetic
interactions among the localized spins.

A serious concern, which is related to the spin-wave spectrum we have obtained,
regards the instability occurring about the center of the Brillouin zone, where 
the spin-wave spectrum becomes purely imaginary.
To overcome this point, one should possibly start from a more complicated 
incommensurate mean-field solution other than the spiral configuration, with 
a lower ground-state energy.
In this way, however, one would unavoidably not obtain a closed-form equation 
for the spin-wave spectrum, since the incommensurability could in general prevent it.
Nonetheless, we expect on physical grounds the spin-wave spectrum we have obtained
with the spiral pattern to survive inclusion of more realistic spin structures,
if one considers only the region close to the boundary of the Brillouin zone,
for which knowledge of the detailed form of the underlying (long-range) spin pattern
appears to be less crucial. 

\section*{Acknowledgments}

We are indebted to P. Salvi for help during the initial stage of this work,
especially in developing  part of the material contained in Appendix C.
We are also indebted to N. Majlis,  W. Brenig, and G. Aeppli for discussions. 

\appendix


\section{Solution of the RPA equations for the dynamical susceptibilities
with an incommensurate spin-spiral ground state}

In this Appendix, we provide the details of the derivation of the explicit form of the
generalized correlation function within RPA and of the associated spin-wave dispersion
relation, which were reported only schematically in Section 2.

We begin by showing how Eq.~(\ref{B-S-RPA}) of the text can be reduced to a closed-form 
equation for the correlation function $\gmathcal{X}$ itself.
To this end, we note from Eq.~(\ref{definition-L}) that the following relations hold 
for $\mu=0$ and $\mu=z$, in the order:

\begin{eqnarray}
\gmathcal{X}_{0,\nu}(x,x') & = & -\frac{i}{4} \sum_{\alpha',\beta'}
\sigma^{\nu}_{\alpha',\beta'} \, L(x+,x'\beta';x^{+}+,x'^{+}\alpha')     \nonumber \\
& - & \frac{i}{4} \sum_{\alpha',\beta'} \sigma^{\nu}_{\alpha',\beta'} \,
L(x-,x'\beta';x^{+}-,x'^{+}\alpha')                                       \label{A-X-rho-nu}
\end{eqnarray}  

\noindent
and                                    

\begin{eqnarray}
\gmathcal{X}_{z,\nu}(x,x') & = & -\frac{i}{4} \sum_{\alpha',\beta'}
\sigma^{\nu}_{\alpha',\beta'} \, L(x+,x'\beta';x^{+}+,x'^{+}\alpha')    \nonumber \\
& + & \frac{i}{4} \sum_{\alpha',\beta'} \sigma^{\nu}_{\alpha',\beta'} \,
L(x-,x'\beta';x^{+}-,x'^{+}\alpha')     \,\, .                              \label{A-X-z-nu}
\end{eqnarray}

\noindent
By adding and subtracting both sides of the above equations, we obtain:

\begin{equation}
\left\{
\begin{array}{ll}
-i \sum_{\alpha',\beta'} \sigma^{\nu}_{\alpha',\beta'} \,
L(x+,x'\beta';x^{+}+,x'^{+}\alpha') = &
2 \, \left[\gmathcal{X}_{0,\nu}(x,x') + \gmathcal{X}_{z,\nu}(x,x') \right]  \\
-i \sum_{\alpha',\beta'} \sigma^{\nu}_{\alpha',\beta'} \,
L(x-,x'\beta';x^{+}-,x'^{+}\alpha') = &
2 \, \left[\gmathcal{X}_{0,\nu}(x,x') - \gmathcal{X}_{z,\nu}(x,x') \right]  \\
\end{array}
\right. \,\, .                                                         \label{A-system-z-nu}
\end{equation}

\noindent
For $\mu=x$ and $\mu=y$ we obtain instead:

\begin{equation}
\left\{
\begin{array}{ll}
-i \sum_{\alpha',\beta'} \sigma^{\nu}_{\alpha',\beta'} \,
L(x-,x'\beta';x^{+}+,x'^{+}\alpha') = &
2 \, \left[\gmathcal{X}_{x,\nu}(x,x') + i\gmathcal{X}_{y,\nu}(x,x') \right]  \\
-i \sum_{\alpha',\beta'} \sigma^{\nu}_{\alpha',\beta'} \,
L(x+,x'\beta';x^{+}-,x'^{+}\alpha') = &
2 \, \left[\gmathcal{X}_{x,\nu}(x,x') - i\gmathcal{X}_{y,\nu}(x,x') \right]  \\
\end{array}
\right.   \,\, ,                                                        \label{A-system-x-y}
\end{equation}

\noindent
with similar relations holding for the non-interacting counterparts of $L$ and
$\gmathcal{X}$.
Entering these relations into Eq.~(\ref{B-S-RPA}), the closed-form expression 
(\ref{integral-equation-0}) results eventually after suitable manipulation.

Next, we specify the form of the non-interacting counterpart $\gmathcal{X}^{(0)}$
of $\gmathcal{X}$, as defined by the first term on the right-hand side of
Eq.~(\ref{B-S-RPA}), namely:

\begin{eqnarray}
\gmathcal{X}^{(0)}_{\mu,\nu}(x,x') & \equiv & -\frac{i}{4} \sum_{\alpha,\beta}
\sum_{\alpha',\beta'} \sigma^{\mu}_{\alpha,\beta} \, \sigma^{\nu}_{\alpha',\beta'} \,
G(x\beta,x'\alpha') \, G(x'\beta',x\alpha)                      \label{A-non-interacting} \\       
& = &\frac{i}{4} \sum_{\alpha,\beta} \sum_{\alpha',\beta'}
\sigma^{\mu}_{\alpha,\beta} \, \sigma^{\nu}_{\alpha',\beta'} \,
\left< T\left[\psi_{\beta}(x) \psi^{\dagger}_{\alpha'}(x')\right] \right>
\left< T\left[\psi_{\beta'}(x')\psi^{\dagger}_{\alpha}(x)\right]
\right> \,\, . \nonumber
\end{eqnarray}

\noindent
To this end, we follow Ref.~\cite{AS-91} and introduce the set of destruction
operators $d_{i\xi}$ along the local spin-quantization axes, which identify the 
spiral pattern of the magnetic ground state within the mean-field approximation.
The field operator (\ref{field-operator}) acquires then the form:

\begin{eqnarray}
\psi_{\alpha}(\gmathbf{r})                                         
& = & \sum_{i} \sum_{\xi} \phi(\gmathbf{r}-\gmathbf{R}_{i}) \,\, 
\gmathcal{R}(\Omega_{i})_{\alpha \xi} \,\,  d_{i \xi}              \label{A-fermion-field} \\
& = & \sum_{i} \sum_{\xi} \sum_{\gmathbf{k}}^{BZ} \phi(\gmathbf{r}-\gmathbf{R}_{i}) \,\,
\gmathcal{R}(\Omega_{i})_{\alpha \xi} \,\,
\frac{\gmathrm{e}^{i\gmathbf{k} \cdot \gmathbf{R}_{i}}}{\sqrt{\gmathcal{N}}} \,\,
 d_{\gmathbf{k} \xi}                                               \nonumber
\end{eqnarray}

\noindent
where $\gmathcal{R}(\Omega_{i})$ is the spin rotation operator associated with the angles
$\Omega_{i}$ and where the Bloch transformation has been introduced as in Section 3, which 
brings the mean-field Hubbard Hamiltonian into block form for each wave vector $\gmathbf{k}$ 
belonging to the Brillouin zone ($BZ$).

Let $W_{\xi,r}(\gmathbf{k})$ (with $\xi=+,-$) be the matrix which diagonalizes the mean-field 
Hubbard Hamiltonian at given $\gmathbf{k}$, such that [cf. Eq.~(\ref{Hamiltonian2})]

\begin{equation}
\sum_{\gmathbf{k}}^{BZ} \sum_{\xi,\xi'} \, d^{\dagger}_{\gmathbf{k} \xi} \,
\gmathcal{H}_{\xi,\xi'}(\gmathbf{k}) \, d_{\gmathbf{k} \xi'} \,             
\, = \sum_{\gmathbf{k}}^{BZ} \sum_{r} \, \gamma^{\dagger}_{\gmathbf{k},r} \, 
\epsilon_{r}(\gmathbf{k}) \, \gamma_{\gmathbf{k},r}               \label{A-gamma-definition-1}
\end{equation}

\noindent
with

\begin{equation}
\gamma_{\gmathbf{k},r} = \sum_{\xi} W^{\dagger}_{r,\xi}(\gmathbf{k}) \,
d_{\gmathbf{k} \xi}                                                \label{A-gamma-definition}
\end{equation}

\noindent
and $\epsilon_{r}(\gmathbf{k})$ given by Eq.~(\ref{eigenvalues}).
Upon averaging over the the broken-symmetry ground state, one then obtains:

\begin{equation}
\left< T[d_{\gmathbf{k} \xi}(t) \, d^{\dagger}_{\gmathbf{k}' \xi'}(t')] \right> 
\, = \sum_{r,r'} W_{\xi,r}(\gmathbf{k}) \, W^{\dagger}_{r',\xi'}(\gmathbf{k}') \,  
\left< T[\gamma_{\gmathbf{k} r}(t) \, \gamma^{\dagger}_{\gmathbf{k}' r'}(t')] \right> 
                                                                        \label{A-d-vs-gamma}
\end{equation}

\noindent
where

\begin{eqnarray}
& & \left< T[\gamma_{\gmathbf{k} r}(t) \, \gamma^{\dagger}_{\gmathbf{k}' r'}(t')] \right>  
                                                               \label{A-time-order-gamma} \\
& & = \delta_{\gmathbf{k},\gmathbf{k}'} \, \delta_{r,r'} \, 
e^{-i \epsilon_{r}(\gmathbf{k})(t-t')} \, 
\left\{\Theta(t-t') \left[ 1 - f_{F}(\epsilon_{r}(\gmathbf{k})) \right] - 
\Theta(t'-t) f_{F}(\epsilon_{r}(\gmathbf{k})) \right\}      \nonumber   
\end{eqnarray}

\noindent
$\Theta(t)$ being the unit step function.
Introducing further the tensor $T(\Omega_{i})$ via the relation

\begin{equation}
\sum_{\alpha,\beta} \gmathcal{R}^{\dagger}(\Omega_{i})_{\xi \alpha} \,
\sigma^{\mu}_{\alpha,\beta} \, \gmathcal{R}(\Omega_{i})_{\beta,\xi'} = 
\sum_{\nu}T_{\mu \nu}(\Omega_i) \, \sigma^{\nu}_{\xi,\xi'}  \,\, ,    \label{A-T-definition}   
\end{equation}

\noindent
such that

\begin{equation}
 T(\Omega_{i}) =
\left(
\begin{array}{cccc}
1 &       0        & 0 &      0        \\
0 &  \cos \theta_i & 0 & \sin \theta_i \\
0 &       0        & 1 &      0        \\
0 & -\sin \theta_i & 0 & \cos \theta_i \\
\end{array}
\right)                                                                  \label{A-T-cos-sin}
\end{equation}

\noindent
where $\theta_{i}=\gmathbf{Q}\cdot\gmathbf{R}_{i}$ within the spiral-spin 
pattern we are considering,\cite{AS-91} and approximating

\begin{equation}
\phi^{*}(\gmathbf{r}-\gmathbf{R}_{i}) \,\, \phi(\gmathbf{r}-\gmathbf{R}_{j}) \, \cong \, 
|\phi(\gmathbf{r}-\gmathbf{R}_{i})|^{2} \,\, \delta_{i,j}                \label{A-non-overlap}
\end{equation}

\noindent
for localized atomic (Wannier) orbitals, we obtain the following expression for 
the frequency Fourier transform of the non-interacting correlation function:

\begin{eqnarray} 
\gmathcal{X}^{(0)}_{\mu,\nu}(\gmathbf{r},\gmathbf{r}';\omega) & = & 
\frac{1}{4\gmathcal{N}^{2}} \sum_{i,j} \sum_{\gmathbf{k},\gmathbf{k}'}^{BZ} \,
e^{i(\gmathbf{k}-\gmathbf{k}')\cdot(\gmathbf{R}_{i}-\gmathbf{R}_{j})} \,
|\phi(\gmathbf{r}-\gmathbf{R}_{i})|^{2} \, |\phi(\gmathbf{r}'-\gmathbf{R}_{j})|^{2} \nonumber \\
& \times & \sum_{r,r'} \sum_{\mu',\nu'} T_{\mu,\mu'}(\Omega_{i}) \,
T_{\nu,\nu'}(\Omega_{j}) \, F^{\mu'}_{r',r}(\gmathbf{k}',\gmathbf{k}) \,
F^{\nu'}_{r,r'}(\gmathbf{k},\gmathbf{k}')                                         \nonumber \\ 
& \times & \gmathcal{F}_{r,r'}(\gmathbf{k},\gmathbf{k}',\omega)  \,\, .    \label{A-Chi-0-r-r'}
\end{eqnarray}

\noindent
Here we have introduced the notation: 

\begin{equation}
F^{\mu}_{r,r'}(\gmathbf{k},\gmathbf{k}') \equiv 
\sum_{\xi,\xi'} W^{\dagger}_{r,\xi}(\gmathbf{k})  \,     
\sigma^{\mu}_{\xi,\xi'} \,  W_{\xi',r'}(\gmathbf{k}')                  \label{A-F-definition}
\end{equation}

\noindent
as well as

\begin{eqnarray}
\gmathcal{F}_{r,r'}(\gmathbf{k},\gmathbf{k}',\omega)  & = &      
\frac{ \left[ 1 - f_{F}(\epsilon_{r}(\gmathbf{k})) \right] 
f_{F}(\epsilon_{r'}(\gmathbf{k}'))}
{\omega - \epsilon_{r}(\gmathbf{k}) + \epsilon_{r'}(\gmathbf{k}') + i\eta}   \nonumber \\
& - & \frac{ \left[ 1 - f_{F}(\epsilon_{r'}(\gmathbf{k}')) \right]
f_{F}(\epsilon_{r}(\gmathbf{k}))}
{\omega - \epsilon_{r}(\gmathbf{k}) + \epsilon_{r'}(\gmathbf{k}')-i\eta}  
                                                                 \label{A-bubble-definition}
\end{eqnarray}

\noindent
where $f_{F}(\epsilon)$ is the Fermi function.
Equation (\ref{FT-chi-O}) of the text is thus recovered.

We have mentioned in Section 2 that the solution to the integral equation 
(\ref{integral-equation-0}) could be considerably simplified, provided the matrix 
$T(\Omega_{i})$ given by Eq.~(\ref{A-T-cos-sin}) were preliminary brought to diagonal 
form by a suitable unitary transformation.
This transformation reads:
 
\begin{equation}
\bar{\sigma}^{a} = \sum_{\mu=(0,x,y,z)} B_{a \mu} \,\sigma^{\mu}    \label{A-transformation}
\end{equation}

\noindent
with $a=(0,1,2,3)$ and where

\begin{equation}
B \, = \,
\left(
\begin{array}{rrrr}
1 &         0          & 0 &         0           \\
0 & \frac{1}{\sqrt{2}} & 0 & \frac{i}{\sqrt{2}}  \\
0 &         0          & 1 &         0           \\
0 & \frac{1}{\sqrt{2}} & 0 & \frac{-i}{\sqrt{2}} 
\end{array}
\right) \,\, .                                                        \label{A-B-definition}
\end{equation}

\noindent
In this way, Eq.(\ref{A-F-definition}) is replaced by:

\begin{equation} 
\bar{F}^{a}_{r,r'}(\gmathbf{k},\gmathbf{k}') = \sum_{\mu=(0,x,y,z)}
B_{a \mu} \, F^{\mu}_{r,r'}(\gmathbf{k},\gmathbf{k}')   \,\, ,       \label{A-Fbar-definition}
\end{equation}

\noindent
and the correlation functions transform according to the rule:

\begin{equation}
\bar{\gmathcal{X}}_{ab} = \sum_{\mu,\nu} B_{a \mu} \, \gmathcal{X}_{\mu \nu} \,
\left( B^{T} \right)_{\nu b}          \,\, .                               \label{A-Chi-bar} 
\end{equation}

\noindent
The second term on the right-hand side of the integral equation (\ref{integral-equation-0}) 
then transforms as follows:

\begin{equation}
B \, \gmathcal{X}^{(0)} \, \epsilon \, \gmathcal{X} \, B^{T} \, = \,
B \, \gmathcal{X}^{(0)} \, B^{T} \, (B^{T})^{-1} \, \epsilon \, 
B^{-1} \, B \, \gmathcal{X} B^{T} \, = \,
\bar{\gmathcal{X}}^{(0)} \, \bar{\epsilon} \, \bar{\gmathcal{X}}     \label{A-matrix-equality}
\end{equation}

\noindent
in matrix notation, where

\begin{equation}
\bar{\epsilon} \, = \, \left( B^{T} \right)^{-1} \, \epsilon \, B^{-1} \, = \,
\left(
\begin{array}{rrrr}
1 &  0 &  0 &  0 \\
0 &  0 &  0 & -1 \\
0 &  0 & -1 &  0 \\
0 & -1 &  0 &  0
\end{array}
\right)       \,\, ;                                                   \label{A-epsilon-bar}
\end{equation}

\noindent
at the same time

\begin{equation}
\bar{T}(\Omega_{i}) = B \, T(\Omega_{i}) \, B^{-1} =
\left(
\begin{array}{cccc}
1 &                0                    & 0 &               0                    \\
0 & e^{-i\gmathbf{Q}\cdot\gmathbf{R}_{i}} & 0 &               0                    \\
0 &                0                    & 1 &               0                    \\
0 &                0                    & 0 & e^{i\gmathbf{Q}\cdot\gmathbf{R}_{i}}
\end{array}
\right)                                                                \label{A-T-bar-equal}
\end{equation}

\noindent
becomes diagonal, as anticipated.
Equation (\ref{T-bar-matrix}) of the text then follows.

There remains to solve Eq.~(\ref{matrix-equation}) of the text explicitly.
To this end, we introduce the compact notation

\begin{equation}
\gamma_{a} =
\left\{
\begin{array}{rcc}
 0 & & (a=0) \\
-1 & & (a=1) \\      
 0 & & (a=2) \\
+1 & & (a=3)
\end{array}
\right.       \,\, ,                                            \label{A-gamma-a-definition}
\end{equation}

\noindent
in such a way that (the lattice Fourier transform of) Eq.~(\ref{A-T-bar-equal}) reads: 

\begin{equation} 
\bar{T}_{a a'}(\gmathbf{k}) \, = \, \delta_{a,a'} \,
\Delta(\gmathbf{k}+\gamma_{a}\gmathbf{Q})                              \label{A- compact-form}        
\end{equation}

\noindent
$\Delta(\gmathbf{k})$ being the lattice Kronecker delta function.
In the transformed basis (cf. Eq.~(\ref{A-Chi-bar})), the non-interacting part 
(\ref{chi-o-cap}) of the correlation function then takes the form: 
 
\begin{equation}
\hat{\gmathcal{X}}^{(0)}_{ab}(\gmathbf{q},\gmathbf{q}';\omega) \, = \,
\Delta(\gmathbf{q}-\gmathbf{q}'-(\gamma_{a}+\gamma_{b})\gmathbf{Q}) \,\,
X^{(0)}_{a b}(\gmathbf{q};\omega|\gmathbf{Q})                              \label{A-rewriting} 
\end{equation}             

\noindent
where we have set

\begin{eqnarray}
X^{(0)}_{a b}(\gmathbf{q};\omega|\gmathbf{Q})  & = &                         
\frac{1}{4\gmathcal{N}} \sum_{\gmathbf{k}}^{BZ} \sum_{r,r'} \,
\bar{F}^{a}_{r',r}(\gmathbf{k}-\gmathbf{q}+\gamma_{a}\gmathbf{Q},\gmathbf{k}) \,\, 
\bar{F}^{b}_{r,r'}(\gmathbf{k},\gmathbf{k}-\gmathbf{q}+\gamma_{a}\gmathbf{Q})       \nonumber \\
& \times & \gmathcal{F}_{r r'}(\gmathbf{k},\gmathbf{k}-\gmathbf{q}+\gamma_{a}\gmathbf{Q}) 
\,\, .                                                                          \label{A-Xo}  
\end{eqnarray}

\noindent
Owing to the wave vector conserving Kronecker delta function in Eq.~(\ref{A-rewriting}), 
the integral equation (\ref{matrix-equation}) becomes:

\begin{eqnarray}
\hat{\gmathcal{X}}_{ab}(\gmathbf{q},\gmathbf{q}';\omega) & = &
\Delta(\gmathbf{q}-\gmathbf{q}'-(\gamma_{a}+\gamma_{b})\gmathbf{Q}) \,
X^{(0)}_{a b}(\gmathbf{q};\omega|\gmathbf{Q})                                             \nonumber \\
& + & 2 \, U \left[ X^{(0)}_{a0}(\gmathbf{q};\omega|\gmathbf{Q}) \,
\hat{\gmathcal{X}}_{0b}(\gmathbf{q}- \gamma_{a} \gmathbf{Q},\gmathbf{q}';\omega) \right.    \nonumber \\    
& - & \left. X^{(0)}_{a1}(\gmathbf{q};\omega|\gmathbf{Q}) \,
\hat{\gmathcal{X}}_{3b}(\gmathbf{q}-(\gamma_{a}-1)\gmathbf{Q},\gmathbf{q}';\omega) \right.  \nonumber \\     
& - & \left. X^{(0)}_{a2}(\gmathbf{q};\omega|\gmathbf{Q}) \, 
\hat{\gmathcal{X}}_{2b}(\gmathbf{q}-\gamma_{a}\gmathbf{Q},\gmathbf{q}';\omega) \right.      \nonumber \\   
& - & \left. X^{(0)}_{a3}(\gmathbf{q};\omega|\gmathbf{Q}) \,
\hat{\gmathcal{X}}_{1b}(\gmathbf{q}-(\gamma_{a}+1) \gmathbf{Q},\gmathbf{q}';\omega) \right]  \,\, .                                                                                                   \label{A-algebraic-X-hat}
\end{eqnarray}

\noindent
Note that the wave vector arguments on the right-hand side of the above expression
depend on the index $a$ of the matrix element.
To avoid this feature, we let $\gmathbf{q} \rightarrow \gmathbf{q} + \gamma_{a} \gmathbf{Q}$
everywhere in the above expression and obtain:

\begin{eqnarray}
\hat{\gmathcal{X}}_{ab}(\gmathbf{q}+\gamma_{a} \gmathbf{Q},\gmathbf{q}';\omega) & = &
\Delta(\gmathbf{q}-\gmathbf{q}'-\gamma_{b}\gmathbf{Q}) \, 
X^{(0)}_{ab}(\gmathbf{q}+\gamma_{a}\gmathbf{Q};\omega|\gmathbf{Q})           \nonumber \\
& + & 2 \, U \left[ X^{(0)}_{a0}(\gmathbf{q}+\gamma_{a}\gmathbf{Q};\omega|\gmathbf{Q}) \,
\hat{\gmathcal{X}}_{0b}(\gmathbf{q},\gmathbf{q}';\omega) \right.             \nonumber \\    
& - & \left. X^{(0)}_{a1}(\gmathbf{q}+\gamma_{a}\gmathbf{Q};\omega|\gmathbf{Q}) \,
\hat{\gmathcal{X}}_{3b}(\gmathbf{q}+\gmathbf{Q},\gmathbf{q}';\omega) \right.  \nonumber \\     
& - & \left. X^{(0)}_{a2}(\gmathbf{q}+\gamma_{a}\gmathbf{Q};\omega|\gmathbf{Q}) \,
\hat{\gmathcal{X}}_{2b}(\gmathbf{q},\gmathbf{q}';\omega) \right.             \nonumber \\   
& - & \left. X^{(0)}_{a3}(\gmathbf{q}+\gamma_{a}\gmathbf{Q};\omega|\gmathbf{Q}) \,
\hat{\gmathcal{X}}_{1b}(\gmathbf{q}-\gmathbf{Q},\gmathbf{Q}';\omega)
\right] \,\, . \nonumber    
\end{eqnarray}

\noindent
For any given value of the index $b$, this relation can then be cast in the
equivalent form:

\begin{equation}
\left[ \gmathbf{1} + X(\gmathbf{q},\omega) \right]
\left(
\begin{array}{c}
\hat{\gmathcal{X}}_{0b}(\gmathbf{q},\gmathbf{q}';\omega)             \\
\hat{\gmathcal{X}}_{1b}(\gmathbf{q}-\gmathbf{Q},\gmathbf{q}';\omega)  \\
\hat{\gmathcal{X}}_{2b}(\gmathbf{q},\gmathbf{q}';\omega)             \\
\hat{\gmathcal{X}}_{3b}(\gmathbf{q}+\gmathbf{Q},\gmathbf{q}';\omega)
\end{array}
\right)
=
\left(
\begin{array}{c}
X^{(0)}_{0b}(\gmathbf{q};\omega|\gmathbf{Q})            \\
X^{(0)}_{1b}(\gmathbf{q}-\gmathbf{Q};\omega|\gmathbf{Q}) \\
X^{(0)}_{2b}(\gmathbf{q};\omega|\gmathbf{Q})            \\
X^{(0)}_{3b}(\gmathbf{q}+\gmathbf{Q};\omega|\gmathbf{Q})
\end{array}
\right)
\Delta(\gmathbf{q}-\gmathbf{q}'-\gamma_{b}\gmathbf{Q})                   \label{A-Chi-hat-vs-X}
\end{equation}

\noindent
where we have set

\begin{eqnarray}
& & X(\gmathbf{q},\omega) \, = \, 2 \, U                                \nonumber \\
& &  \times \left(
\begin{array}{cccc}
-X^{(0)}_{00}(\gmathbf{q};\omega|\gmathbf{Q})              &  X^{(0)}_{03}(\gmathbf{q};\omega|\gmathbf{Q})              &
 X^{(0)}_{02}(\gmathbf{q};\omega|\gmathbf{Q})              &  X^{(0)}_{01}(\gmathbf{q};\omega|\gmathbf{Q})              \\
-X^{(0)}_{10}(\gmathbf{q}-\gmathbf{Q};\omega|\gmathbf{Q})   &  X^{(0)}_{13}(\gmathbf{q}-\gmathbf{Q};\omega|\gmathbf{Q})   &
 X^{(0)}_{12}(\gmathbf{q}-\gmathbf{Q};\omega|\gmathbf{Q})   &  X^{(0)}_{11}(\gmathbf{q}-\gmathbf{Q};\omega|\gmathbf{Q})   \\
-X^{(0)}_{20}(\gmathbf{q};\omega|\gmathbf{Q})              &  X^{(0)}_{23}(\gmathbf{Q};\omega|\gmathbf{Q})              &
 X^{(0)}_{22}(\gmathbf{q};\omega|\gmathbf{Q})              &  X^{(0)}_{21}(\gmathbf{Q};\omega|\gmathbf{Q})              \\
-X^{(0)}_{30}(\gmathbf{q}+\gmathbf{Q};\omega|\gmathbf{Q})   &  X^{(0)}_{33}(\gmathbf{q}+\gmathbf{Q};\omega|\gmathbf{Q})   &
 X^{(0)}_{32}(\gmathbf{q}+\gmathbf{Q};\omega|\gmathbf{Q})   &  X^{(0)}_{31}(\gmathbf{Q}+\gmathbf{Q};\omega|\gmathbf{Q})   \\
\end{array}
\right)    \,\, .                                                     \nonumber \\
                                                                                                 \label{A-big-matrix}
\end{eqnarray}

\noindent
Solving for $\hat{\gmathcal{X}}_{ab}$ in Eq.~(\ref{A-Chi-hat-vs-X}), we obtain eventually:

\begin{eqnarray}
\hat{\gmathcal{X}}_{ab}(\gmathbf{q}+\gamma_{a}\gmathbf{Q},\gmathbf{q'};\omega) & = &
\sum_{a'} \, [\gmathbf{1} + X(\gmathbf{q},\omega)]_{aa'}^{-1}         \nonumber  \\ 
& \times & X^{(0)}_{a'b}(\gmathbf{q}+\gamma_{a'}\gmathbf{Q};\omega|\gmathbf{Q}) \,
\Delta(\gmathbf{q}-\gmathbf{q'}-\gamma_{b}\gmathbf{Q})                  \label{A-Chi-hat-final}
\end{eqnarray}

\noindent
which coincides with Eq.~(\ref{xxx}) of the text.


\section{Details of the {$\lowercase{t}/U$} expansion}

In this Appendix, we provide details of the $t/U$ expansion of the mean-field
parameters and of the matrix elements of the correlation function which are
necessary to obtain the spin-wave dispersion relation.

We begin by considering the leading ($t=0$) term of the expansion (\ref{epsilon-2}) 
for the mean-field (band) eigenvalues, corresponding to completely flat bands, which
is given by [cf. Eq.~(\ref{eigenvalues})]

\begin{equation}
\epsilon_{r}(\gmathbf{k}) \, = \, U \left(\epsilon_{r}^{(0)}-\mu^{(0)}\right)
\,  = \, U(m_{1} + (-1)^{r} m_{2}) - \mu_{0}  
                                                            \label{B-development-t=0}
\end{equation}

\noindent
with $r=1,2$ and where $\mu_{0}=U \mu^{(0)}$ is the chemical potential at the lowest 
order in $t/U$.  
Since at this order the band eigenvalues do not depend on the wave vector, there is 
no value of the chemical potential $\mu^{(0)}$ satisfying equation (\ref{m1}) at zero 
temperature  and noninteger doping $\delta$.
More precisely, the doping jumps from $\delta=-1$ when $\mu^{(0)}<\epsilon_1^{(0)}$, 
to $\delta=0$ when $\epsilon_1^{(0)}<\mu^{(0)}<\epsilon_2^{(0)}$, and to $\delta=+1$ 
when $\mu^{(0)}>\epsilon_2^{(0)}$.
In order  to consider a continuously doped system, it is therefore necessary to include 
at the outset the next-to-leading order in the expansion (\ref{epsilon-2}) of the band 
eigenvalues, which introduces a band dispersion. 

For definiteness, we shall consider the case $\delta>0$  from now on. 
(The case  $\delta<0$ can be recovered by exploiting particle-hole symmetry.) 
In the case $\delta>0$, the lowest band is always filled, and we will consistently use
(the zero-temperature value) $f_{F}(\epsilon_1(\gmathbf{k}))=1$  throughout.  
On the other hand, the zeroth order chemical potential for $0<\delta<1$ must be chosen as
$\mu^{(0)}=\epsilon_{2}^{(0)} $, and the Fermi function in the upper
band can be expanded as

\begin{eqnarray}
& & f_{F}(\epsilon_{2}(\gmathbf{k})) \, = \, 1 - \Theta 
\left(
\epsilon_{2}^{(1)}(\gmathbf{k})-\mu^{(1)} \right)                           \label{B-Fermi-2} \\
& - & \delta \left(
\epsilon_{2}^{(1)}(\gmathbf{k})-\mu^{(1)} \right) \,                     
\left[\left(\frac{t}{U}\right)(\epsilon_{2}^{(2)}(\gmathbf{k})-\mu^{(2)})  + 
\left(\frac{t}{U}\right)^2(\epsilon_{2}^{(3)}(\gmathbf{k})-\mu^{(3)}) 
\right]                                                                         \nonumber \\
& - & \frac{1}{2} \, \delta' \left(
\epsilon_{2}^{(1)}(\gmathbf{k})-\mu^{(1)} \right)  \,
\left(\frac{t}{U}\right)^{2}(\epsilon_{2}^{(2)}(\gmathbf{k})-\mu^{(2)})^{2}    
\, + \, \cdots                     \, ,                                          \nonumber     
\end{eqnarray}

\noindent
where we have considered the zero-temperature limit of the Fermi function
and of its derivatives, and introduced  the step function $\Theta $ as well as the Dirac  
$\delta(x)$ function (not to be confused with doping).
In this way, Eq.~(\ref{m1}) can be expressed in powers of $t/U$, yielding

\begin{eqnarray}
&& \frac{1}{\gmathcal{N}} \sum_{\gmathbf{k}}^{BZ}  \,
\Theta\left( 
\epsilon_{2}^{(1)}(\gmathbf{k})-\mu^{(1)}\right) \, = \, 1 - \delta \, ,   \nonumber \\ 
&&  \frac{1}{\gmathcal{N}} \sum_{\gmathbf{k}}^{BZ}  \,
\delta\left(
\epsilon_{2}^{(1)}(\gmathbf{k})-\mu^{(1)}\right)
(\epsilon_{2}^{(2)}(\gmathbf{k})-\mu^{(2)}) \, = \, 0  \, ,                \nonumber \\
&& \frac{1}{2\gmathcal{N}} \sum_{\gmathbf{k}}^{BZ} \, \left[
\delta' \left(
\epsilon_{2}^{(1)}(\gmathbf{k})-\mu^{(1)}\right)
(\epsilon_{2}^{(2)}(\gmathbf{k})-\mu^{(2)})^{2} 
+ 2 \  \delta\ \left(
\epsilon_{2}^{(1)}(\gmathbf{k})-\mu^{(1)}\right)
(\epsilon_{2}^{(3)}(\gmathbf{k})-\mu^{(3)})
\right]
\, = \, 0 \, .                                                            \nonumber \\
                                                                                \label{B-a1}
\end{eqnarray}

\noindent
Similarly, the small $t/U$ expansion (\ref{m2-2}) of Eq.~(\ref{m2}) yields 
[cf. also Eqs.~(\ref{W1})-(\ref{normalization})]:

\begin{eqnarray}
&& m_{2}^{(0)} + \left(\frac{t}{U}\right) m_{2}^{(1)} +  
\left(\frac{t}{U}\right)^{2} m_{2}^{(2)}  \, + \, \cdots                  \label{B-m-2-0} \\
&& = \frac{1}{2\gmathcal{N}} \sum_{\gmathbf{k}}^{BZ} 
\left[ 1 - 2 \left(\frac{t}{U}\right)^{2} 
\left(\frac{T_{o}(\gmathbf{k})}{2 m_{2}^{(0)}}\right)^{2} \right]                                                    
\left( 1 - 
                             f_{F}(\epsilon_{2}(\gmathbf{k})) \right) + 
\gmathcal{O}\left( \left(\frac{U}{t}\right)^{3} \right) \,\, ,           \nonumber
\end{eqnarray} 

\noindent
where the Fermi function should be again expanded in the form (\ref{B-Fermi-2}). 
One obtains eventually:

\begin{eqnarray}
m_{2}^{(0)} & = & 
\frac{1}{2\gmathcal{N}} \sum_{\gmathbf{k}}^{BZ}  \,
 \Theta \left( 
\epsilon_{2}^{(1)}(\gmathbf{k})-\mu^{(1)} \right)  \, ,                 \nonumber \\ 
m_{2}^{(1)} & = & \frac{1}{2\gmathcal{N}} \sum_{\gmathbf{k}}^{BZ}  \,
 \delta \left( 
\epsilon_{2}^{(1)}(\gmathbf{k})-\mu^{(1)} \right)
(\epsilon_{2}^{(2)}(\gmathbf{k})-\mu^{(2)})  \, ,                       \nonumber \\ 
m_{2}^{(2)} & = & 
- \, \frac{1}{\gmathcal{N}} \sum_{\gmathbf{k}}^{BZ} \,   
 \left(\frac{T_{o}(\gmathbf{k})}{2 m_{2}^{(0)}}\right)^{2}
\Theta\left(
\epsilon_{2}^{(1)}(\gmathbf{k})-\mu^{(1)}\right)                        \nonumber \\
& + & \frac{1}{4\gmathcal{N}} \sum_{\gmathbf{k}}^{BZ}  \,
\delta' \left(
\epsilon_{2}^{(1)}(\gmathbf{k})-\mu^{(1)}\right)
(\epsilon_{2}^{(2)}(\gmathbf{k})-\mu^{(2)})^{2}                         \nonumber \\
&+&
\frac{1}{2\gmathcal{N}} \sum_{\gmathbf{k}}^{BZ}  \,
 \delta \left( 
\epsilon_{2}^{(1)}(\gmathbf{k})-\mu^{(1)} \right)
(\epsilon_{2}^{(3)}(\gmathbf{k})-\mu^{(3)})                             \nonumber \\
&=& 
- \, \frac{1}{\gmathcal{N}} \sum_{\gmathbf{k}}^{BZ} \,   
 \left(\frac{T_{o}(\gmathbf{k})}{2 m_{2}^{(0)}}\right)^{2}
 \Theta\left(
\epsilon_{2}^{(1)}(\gmathbf{k})-\mu^{(1)}\right)   \, ,                 \nonumber \\
                                                                                \label{B-a2}
\end{eqnarray}
where in the last step we have used the last of Eqs. (\ref{B-a1}).

There remains to expand in powers of $t/U$ the last the self-consistency equation
(\ref{consistency3}).
For the diagonal spin-spiral solution we obtain:

\begin{eqnarray}
& & 0 \, = \, \sin (Q/2) \, \frac{1}{\gmathcal{N}} \sum_{\gmathbf{k}}^{BZ} \,
(\cos k_{x}+\cos k_{y})
\left( 1+
                             f_{F}(\epsilon_{2}(\gmathbf{k})) \right)         \label{B-Q} \\
& & + \, \left(\frac{t}{U}\right) \cos (Q/2) \frac{1}{m_{2}^{(0)}}
\frac{1}{\gmathcal{N}} \sum_{\gmathbf{k}}^{BZ} \, T_{o}(\gmathbf{k}) \, (\sin k_{x}+\sin k_{y})
\left( 1-
                             f_{F}(\epsilon_{2}(\gmathbf{k})) \right)         \nonumber 
\end{eqnarray}

\noindent
where the expansion (\ref{B-Fermi-2}) has still to be inserted.

From Eq.~(\ref{eigenvalues}),  the argument of the $\Theta$ function is
$T_{e}(\gmathbf{k}) + m_{2}^{(1)} - \mu^{(1)}$ (with 
$T_{e}(\gmathbf{k}) = 2 \, \cos (Q/2) \, (\cos k_{x} + \cos k_{y})$ for
the diagonal spiral
solution), which can take positive as well as negative values.

At the various orders in $t/U$ we then obtain from Eqs.~(\ref{B-a1}) and (\ref{B-a2}):
  
\noindent
(i) order $(t/U)^{0}$

\begin{eqnarray}
& & \frac{1}{\gmathcal{N}} \sum_{\gmathbf{k}}^{BZ} \, \left[ 1 - \Theta 
\left( T_{e}(\gmathbf{k}) + m_{2}^{(1)} - \mu^{(1)} \right) \right] \,
= \, \delta \, ,                                                     \nonumber \\        
&  & m_{2}^{(0)} = \frac{1}{2\gmathcal{N}} \sum_{\gmathbf{k}}^{BZ} \, \Theta
\left( T_{e}(\gmathbf{k}) + m_{2}^{(1)} - \mu^{(1)} \right) \, ;            \label{B-order-0}     
\end{eqnarray}
 
\noindent
(ii)  order $(t/U)^{1}$ 

\begin{eqnarray}
& & \frac{1}{\gmathcal{N}} \sum_{\gmathbf{k}}^{BZ} \, \delta 
\left( T_{e}(\gmathbf{k}) + m_{2}^{(1)} - \mu^{(1)} \right) \,
(\epsilon_{2}^{(2)}(\gmathbf{k})-\mu^{(2)}) =0\ ,                     \nonumber \\
& & m_{2}^{(1)} = \frac{1}{2\gmathcal{N}} \sum_{\gmathbf{k}}^{BZ} \, \delta
\left( T_{e}(\gmathbf{k}) + m_{2}^{(1)} - \mu^{(1)} \right) \,
(\epsilon_{2}^{(2)}(\gmathbf{k})-\mu^{(2)}) \, ;                            \label{B-order-1}      
\end{eqnarray}

\noindent
(iii)  order $(t/U)^{2}$
 
\begin{equation}
 m_{2}^{(2)} = - \frac{1}{\gmathcal{N}} \sum_{\gmathbf{k}}^{BZ}  
\left( \frac{T_{o}(\gmathbf{k})}{2m_{2}^{(0)}} \right)^{2} \, \Theta
\left( T_{e}(\gmathbf{k}) + m_{2}^{(1)} - \mu^{(1)} \right)  \, ,           \label{B-order-2}
\end{equation}
and we don't need to evaluate $(\epsilon_{2}^{(3)}(\gmathbf{k})-\mu^{(3)})$
at the order we are considering.
It is convenient to solve these equations order-by-order for the two
variables $\mu^{(i)} - m_2^{(i)}$, and $m_2^{(i)}$ ($i=0,1,2$).

Using at this point the method developed in Appendix C for performing the $\gmathbf{k}$
summation over the relevant portions of the Brillouin zone, we obtain 
eventually the following results for the mean-field parameters at the leading
orders in $\delta$:

\begin{equation}
\left\{
\begin{array}{rcl}                               
m_2^{(0)} & = & (1-\delta)/2              \nonumber \\
m_2^{(1)} & = & 0                         \nonumber \\
m_2^{(2)} & = & - 4 \, \sin ^2(Q/2) + \gmathcal{O}(\delta) \,  ,
\end{array}
\right.                                   \label{B-m}
\end{equation}

\noindent
and

\begin{equation}
\left\{
\begin{array}{rcl}
\mu^{(0)} - m_{2}^{(0)} & = & m_{1} = (1+\delta)/2              \nonumber \\
\mu^{(1)} - m_{2}^{(1)} & = &
4 \, \cos(Q/2) \, (1 - \pi \delta + \gmathcal{O}(\delta^{2}))    \nonumber \\
\mu^{(2)} - m_{2}^{(2)} & = & 0 + \gmathcal{O}(\delta) \,\, .
\end{array}
\right.                                                                         \label{B-mm}
\end{equation}

There remains to determine the magnitude $Q$ of the characteristic wave vector
from the self-consistency condition (\ref{B-Q}), where for the Fermi function we
use the expansion (\ref{B-Fermi-2}).
Equation (\ref{B-Q}) then becomes:

\begin{eqnarray}
& & \left(\frac{t}{U}\right) \sin (Q/2) \, \frac{1}{\gmathcal{N}} \sum_{\gmathbf{k}}^{BZ} \,
(\cos k_{x}+\cos k_{y}) \, \left[ \phantom{\frac{L}{R}} 2 -  \Theta
\left( T_{e}(\gmathbf{k}) + m_{2}^{(1)} - \mu^{(1)} \right) \right.    \nonumber  \\
& & \left. - \left(\frac{t}{U}\right) \delta 
\left( T_{e}(\gmathbf{k}) + m_{2}^{(1)} - \mu^{(1)} \right) \,
\left( \epsilon_{2}^{(2)} (\gmathbf{k}) - \mu^{(2)} \right) \right]
+ 2 \, \left(\frac{t}{U}\right)^{2} \, \cos (Q/2)                     \nonumber  \\
& & \times \frac{1}{2 m_{2}^{(0)}} \frac{1}{\gmathcal{N}} \sum_{\gmathbf{k}}^{BZ} \, 
(\sin k_{x}+\sin k_{y}) \, T_{o}(\gmathbf{k}) \, \Theta
\left( T_{e}(\gmathbf{k}) + m_{2}^{(1)} - \mu^{(1)} \right)            \nonumber  \\      
& & = 0 \,\, .                                                                \label{B-long}
\end{eqnarray}

\noindent
In particular, at the lowest order in $t/U$ we obtain:

\begin{equation}
\sin (Q/2) \, (2 \, \delta + \gmathcal{O}(\delta^{2})) \, = \, 0  \,\, .     \label{B-zero-1}
\end{equation}

\noindent
For fixed $\delta$ and large $U/t$, one thus has only 
 the \emph{ferromagnetic\/} solution $Q=0$. 
However, if one allows $\delta$ to be of the order $t/U$, 
Eq.(\ref{B-zero-1}) has to be considered together with the next term
in $t/U$, and we obtain instead:

\begin{equation}
\sin (Q/2) \, (2 \, \delta + \gmathcal{O}(\delta^{2})) +
4 \, \sin (Q/2) \, \cos(Q/2) \, (1 + \gmathcal{O}(\delta)) \, = \, 0         \label{B-zero-2}
\end{equation}

\noindent
which yields the two solutions

\[ \begin{array}{cc}
\sin (Q/2) \, = \, 0                       &   \mbox{ (ferromagnet) }     \\               
\cos (Q/2) \, = \, - \delta U/2t           &   \mbox{ (diagonal spiral) }  \,\, ,   
\end{array} \]

\noindent
with the spiral solution being energetically favored over the ferromagnetic 
solution whenever it exists.
From the above equation, it is clear that  the spiral solution exists for any $\delta \leq
2t/U$, which is consistent with our assumption that the doping parameter $\delta$ is at 
most of order $t/U$ in the spiral phase, thus justifying the expansion of Appendix~C. 
The transition to the ferromagnetic state is second order, as the
incommensurability vector $Q$ 
decreases continuously from $Q=\pi$ to $Q=2\pi$ with increasing $\delta$. 

We pass finally to consider the $t/U$ expansion of the matrix elements of the
correlation function.
To begin with, we use the expressions (\ref{W1})-(\ref{normalization}) for
the eigenvectors of the mean-field Hamiltonian and obtain from the
definitions (\ref{da}) and (\ref{du}):

\begin{eqnarray}
\gmathcal{X}_{0}^{+,-}(\gmathbf{q},\omega) & = & \frac{1}{2\gmathcal{N}} \sum_{\gmathbf{k}}^{BZ} \,
\left\{ \phantom{\frac{L}{R}} \left[1- \left(\frac{t}{U}\right)^{2} 
\left( T_{o}^{2}(\gmathbf{k}) + T_{o}^{2}(\gmathbf{k}-\gmathbf{q}) \right) \right]
\gmathcal{F}_{2,1}(\gmathbf{k},\gmathbf{k}-\gmathbf{q},\omega) \right.                 \nonumber \\
& + & \left. \left(\frac{t}{U}\right)^{2} 
\left[ T_{o}^{2}(\gmathbf{k}) \gmathcal{F}_{1,1}(\gmathbf{k},\gmathbf{k}-\gmathbf{q},\omega) + 
T_{o}^{2}(\gmathbf{k}-\gmathbf{q}) \gmathcal{F}_{2,2}(\gmathbf{k},\gmathbf{k}-\gmathbf{q},\omega) \right]
\phantom{\frac{L}{R}} \right\}                                                     \nonumber \\
& + & \gmathcal{O}\left(\left(t/U\right)^{3}\right)                        \label{plus-minus}
\end{eqnarray}        

\begin{eqnarray}             
\gmathcal{X}_{0}^{+,+}(\gmathbf{q},\omega) & = & \frac{1}{2\gmathcal{N}} \sum_{\gmathbf{k}}^{BZ} \,
\left\{ \left(\frac{t}{U}\right)^{2} T_{o}(\gmathbf{k}) \, T_{o}(\gmathbf{k}-\gmathbf{q}) \right.  \nonumber \\
& + & \left. \left[ \gmathcal{F}_{1,2}(\gmathbf{k},\gmathbf{k}-\gmathbf{q},\omega) +  
\gmathcal{F}_{2,1}(\gmathbf{k},\gmathbf{k}-\gmathbf{q},\omega) \right. \right.                      \nonumber \\
& - & \left.\left. \gmathcal{F}_{1,1}(\gmathbf{k},\gmathbf{k}-\gmathbf{q},\omega) -       
\gmathcal{F}_{2,2}(\gmathbf{k},\gmathbf{k}-\gmathbf{q},\omega) \right]
\phantom{\frac{L}{R}} \right\} + \gmathcal{O}\left(\left(t/U\right)^{3}\right)                   \nonumber \\
                                                                                                \label{plus-plus}
\end{eqnarray}

\noindent
for the matrix elements entering the final expression (\ref{omega-2}) of the spin-wave 
dispersion relation.
For the remaining matrix elements we obtain instead:

\begin{eqnarray}
\gmathcal{X}_{0}^{0,0}(\gmathbf{q},\omega) & = & \frac{1}{4\gmathcal{N}} \sum_{\gmathbf{k}}^{BZ} \,
\left\{ \left(\frac{t}{U}\right)^{2} 
\left( T_{o}(\gmathbf{k}) - T_{o}(\gmathbf{k}-\gmathbf{q}) \right)^{2} \right.           \nonumber \\
& \times & \left. \left[ \gmathcal{F}_{1,2}(\gmathbf{k},\gmathbf{k}-\gmathbf{q},\omega) +
\gmathcal{F}_{2,1}(\gmathbf{k},\gmathbf{k}-\gmathbf{q},\omega) \right] \right.            \nonumber \\
& + & \left. \left[ 1- \left(\frac{t}{U}\right)^{2} 
\left( T_{o}(\gmathbf{k}) - T_{o}(\gmathbf{k}-\gmathbf{q}) \right)^{2} \right] \right.   \nonumber \\
& \times & \left. \left[ \gmathcal{F}_{1,1}(\gmathbf{k},\gmathbf{k}-\gmathbf{q},\omega) +
                         \gmathcal{F}_{2,2}(\gmathbf{k},\gmathbf{k}-\gmathbf{q},\omega) \right]
\phantom{\frac{L}{R}} \right\} + \gmathcal{O}\left(\left(t/U\right)^{3}\right) \,\, ,  \nonumber \\
                                                                             \label{rho,rho}
\end{eqnarray}

\begin{eqnarray}
\gmathcal{X}_{0}^{0,-}(\gmathbf{q},\omega) & = & \frac{1}{4\gmathcal{N}} \sum_{\gmathbf{k}}^{BZ} \, i \sqrt{2} 
\, \left\{ \left(\frac{t}{U}\right) \frac{1}{1-\delta} \,    
\left( T_{o}(\gmathbf{k}-\gmathbf{q}) - T_{o}(\gmathbf{k}) \right) \,
\gmathcal{F}_{2,1}(\gmathbf{k},\gmathbf{k}-\gmathbf{q},\omega) \right.                    \nonumber \\
& + & \left. \left(\frac{t}{U}\right) \frac{1}{1-\delta} \, \left[ 
T_{o}(\gmathbf{k}) \, \gmathcal{F}_{1,1}(\gmathbf{k},\gmathbf{k}-\gmathbf{q},\omega) - 
T_{o}(\gmathbf{k}-\gmathbf{q}) \,\gmathcal{F}_{2,2}(\gmathbf{k},\gmathbf{k}-\gmathbf{q},\omega)   
\right] \right\}                                                                      \nonumber \\
& + & \gmathcal{O}\left(\left(t/U\right)^{3}\right) \,\, ,                  \label{rho,minus}
\end{eqnarray}

\begin{eqnarray}
\gmathcal{X}_{0}^{0,z}(\gmathbf{q},\omega) & = & \frac{1}{4\gmathcal{N}} \sum_{\gmathbf{k}}^{BZ} \,
\left\{ \left(\frac{t}{U}\right)^{2} 
\left(T_{o}^{2}(\gmathbf{k}) - T_{o}^{2}(\gmathbf{k}-\gmathbf{q})\right) \right.         \nonumber \\
& \times & \left. \left[   \gmathcal{F}_{2,1}(\gmathbf{k},\gmathbf{k}-\gmathbf{q},\omega) -
\gmathcal{F}_{1,2}(\gmathbf{k},\gmathbf{k}-\gmathbf{q},\omega) \right] \right.            \nonumber \\
& + & \left. \left[ 1- \left(\frac{t}{U}\right)^{2} 
\left( T_{o}^{2}(\gmathbf{k}) + T_{o}^{2}(\gmathbf{k}-\gmathbf{q}) \right)
\right] \right.                                                                       \nonumber \\
& \times & \left. \left[ \gmathcal{F}_{1,1}(\gmathbf{k},\gmathbf{k}-\gmathbf{q},\omega) -
\gmathcal{F}_{2,2}(\gmathbf{k},\gmathbf{k}-\gmathbf{q},\omega) \right] \phantom{\frac{L}{R}} \right\}
+ \gmathcal{O}\left(\left(t/U\right)^{3}\right)  \,\, ,                                \nonumber \\
                                                                            \label{rho,zeta}
\end{eqnarray}

\begin{eqnarray}
\gmathcal{X}_{0}^{+,z}(\gmathbf{q},\omega) & = & \frac{1}{4\gmathcal{N}} \sum_{\gmathbf{k}}^{BZ} \, i \sqrt{2}                     
\left\{ \left(\frac{t}{U}\right) \frac{1}{1-\delta} \,
\left( T_{o}(\gmathbf{k}-\gmathbf{q}) + T_{o}(\gmathbf{k}) \right) \,
\gmathcal{F}_{2,1}(\gmathbf{k},\gmathbf{k}-\gmathbf{q},\omega) \right.                   \nonumber \\
& + & \left. \left(\frac{t}{U}\right) \frac{1}{1-\delta} \,
\left[ - T_{o}(\gmathbf{k}) \, \gmathcal{F}_{1,1}(\gmathbf{k},\gmathbf{k}-\gmathbf{q},\omega) -
T_{o}(\gmathbf{k}-\gmathbf{q}) \, \gmathcal{F}_{2,2}(\gmathbf{k},\gmathbf{k}-\gmathbf{q},\omega) 
\right] \right\}                                                                     \nonumber \\
& + & \gmathcal{O}\left(\left(t/U\right)^{3}\right)  \,\, ,                  \label{rho,plus}
\end{eqnarray}

\begin{eqnarray}
\gmathcal{X}_{0}^{z,z}(\gmathbf{q},\omega) & = & \frac{1}{4\gmathcal{N}} \sum_{\gmathbf{k}}^{BZ} \,
\left\{ \left(\frac{t}{U}\right)^{2} 
\left(T_{o}(\gmathbf{k}) + T_{o}(\gmathbf{k}-\gmathbf{q})\right)^{2} \right.             \nonumber \\
& \times &  \left. \left[ \gmathcal{F}_{1,2}(\gmathbf{k},\gmathbf{k}-\gmathbf{q},\omega) +
\gmathcal{F}_{2,1}(\gmathbf{k},\gmathbf{k}-\gmathbf{q},\omega) \right] \right.            \nonumber \\
& + & \left. \left[ 1 - \left(\frac{t}{U}\right)^{2} 
\left(T_{o}(\gmathbf{k})+T_{o}(\gmathbf{k}-\gmathbf{q})\right)^{2} \right] \right. 
                                                                           \label{zeta-zeta} \\
& \times & \left. \left[ \gmathcal{F}_{1,1}(\gmathbf{k},\gmathbf{k}-\gmathbf{q},\omega) +
\gmathcal{F}_{2,2}(\gmathbf{k},\gmathbf{k}\gmathbf{k}-\gmathbf{q},\omega) \right]
\phantom{\frac{L}{R}}          \right\}
+ \gmathcal{O}\left(\left(t/U\right)^{3}\right)   \,\, .                               \nonumber 
\end{eqnarray}

\noindent
Note that in the above expressions  $\gmathcal{F}_{1,1}(\gmathbf{k},\gmathbf{k}-\gmathbf{q},\omega) = 0$
when the doping parameter $\delta \geq 0$.
In addition, $\gmathcal{F}_{1,2}$ can be obtained from $\gmathcal{F}_{2,1}$ through the symmetry 
condition (\ref{symmetry-F}).
Making use of the expansion (\ref{epsilon-2}) for the eigenvalues of the mean-field Hamiltonian,
we write further

\begin{eqnarray}
\gmathcal{F}_{2,2}(\gmathbf{k},\gmathbf{k-q}) & = & \frac{1}{t} \left[f_{F}(\epsilon_{2}(\gmathbf{k-q}))
-f_{F}(\epsilon_{2}(\gmathbf{k}))\right]  \frac{1}{T_{e}(\gmathbf{k-q})-T_{e}(\gmathbf{k})}     \nonumber \\
& + & \gmathcal{O}\left(1/U\right)                                                            \label{intra}
\end{eqnarray}

\noindent
as well as                                          
     
\begin{eqnarray}
\gmathcal{F}_{2,1}(\gmathbf{k},\gmathbf{k-q}) & = & 
\left[f_{F}(\epsilon_{2}(\gmathbf{k}))-1\right] \frac{1}{U}          
\left\{ \frac{1}{(1-\delta-\tilde{\omega})} \right.                                            \nonumber \\
& - &\left. \left(\frac{t}{U}\right) \frac{1}{(1-\delta-\tilde{\omega})^{2}}
        \left[ T_{e}(\gmathbf{k})-T_{e}(\gmathbf{k-q}) \right] +
\left(\frac{t}{U}\right)^{2}   \left[ \frac{1}{(1-\delta-\tilde{\omega})^{2}} \right. \right.  \nonumber \\
& \times & \left. \left. \left(- T_{o}^{2}(\gmathbf{k}) -  T_{o}^{2}(\gmathbf{k-q}) - 2m^{(2)}_{2}\right)    
\right.\right.                                                                                 \nonumber \\
& + & \left. \left. \frac{1}{(1-\delta-\tilde{\omega})^{3}}   
\left( T_{e}(\gmathbf{k}) - T_{e}(\gmathbf{k-q}) \right)^{2} \right] \right\}                    \nonumber \\
& + & \gmathcal{O}\left(t^{3}/U^{4}\right)                                                      \label{inter}
\end{eqnarray}

\noindent
with $\tilde{\omega} = \omega/U$, where the expansion (\ref{B-Fermi-2}) of the Fermi function 
has still to be inserted.


\section{ Sums over the two-dimensional BZ in the small $\delta$ limit}

In this Appendix, we develop a method suitable for performing the $\gmathbf{k}$ summation 
over the relevant portions of the Brillouin zone in powers of the doping parameter $\delta$.
This is, in turn, justified by the fact that $\delta$ in the spiral phase is of order $t/U$,
as shown in Appendix~B.

The typical integral to be evaluated is of the form:

\begin{equation}
\gmathcal{I}(\gamma) \, = \, \frac{1}{\gmathcal{N}} \sum_{\gmathbf{k}}^{BZ} \, g(\gmathbf{k}) 
\, \Theta \left( T_{e}(\gmathbf{k}) + \gamma \right)               \label{typical-integral-0}     
\end{equation}

\noindent
where $T_{e}(\gmathbf{k}) = 2 \cos (Q/2) (\cos k_{x} + \cos k_{y})$ for the diagonal spiral 
solution we are considering  and $\gamma$ is a parameter (which depends on the chemical 
potential $\mu^{(1)}$ at the order ${\gmathcal{O}\left[(t/U)^{0}\right]})$ that in the 
following we shall simply call $\mu$ for simplicity.
Introducing the notation $f_{\mu}$ such that $\gamma = 4 \cos (Q/2) (1 - f_{\mu}/4)$ for
any given $Q$ value,  recalling the definition of the $\Theta$
function, and considering that $\cos (Q/2)<0$ in our solution, we set further:

\begin{equation}
\gmathcal{I}(\gamma) \, = \, \frac{1}{\gmathcal{N}} \sum_{\cos k_{x} + \cos k_{y} 
\leq 2 - f_{\mu}/2} \, g(\gmathbf{k}) \, .                         \label{typical-integral-1}
\end{equation}  

Quite generally, we can introduce a polar representation for the two-dimensional 
wave vector $\gmathbf{k}$ and determine for each value of the polar angle $\phi$
the magnitude $k(\phi)$, such that the equality 

\begin{equation} 
\cos \left( k(\phi) \cos \phi \right) + \cos \left( k(\phi) \sin \phi \right) 
\, = \, 2 - \frac{f_{\mu}}{2}                                                 \label{circle}
\end{equation}

\noindent
is satisfied.
In this way we rewrite the integral (\ref{typical-integral-1}) as follows: 

\begin{equation}
\gmathcal{I}(\gamma) \, = \, \frac{1}{4\pi^{2}} \int_{BZ} \! dk_{x} dk_{y} \, 
g(k_{x},k_{y}) - \frac{1}{4\pi^{2}} \int_{0}^{2\pi} \! d\phi 
\int_{0}^{k(\phi)} \! dk \, k \, g(k,\phi)    \,\, ,                  \label{polar-integral}     
\end{equation} 

\noindent
with a slight (albeit harmless) abuse of notation for the function $g$.

\noindent
In practice, when $f_{\mu}$ is small compared to unity (as it is the case for small 
$\delta$) it is convenient to determine $k(\phi)$ in powers of $f_{\mu}$ by expanding
$\cos \left( k(\phi) \cos \phi \right) + \cos \left( k(\phi) \sin \phi \right)$ in
a Taylor series about $k=0$.
For instance, at the lowest order we obtain from Eq.~(\ref{circle}):

\begin{equation}
k(\phi)^{2} \, + \, \gmathcal{O}\left(k(\phi)^{4}\right) \, = \, f_{\mu}  \label{first-order}
\end{equation}

\noindent
which gives $k(\phi)^{2}=f_{\mu}+\gmathcal{O}\left(f_{\mu}^{2}\right)$.
At the next significant order we obtain instead:

\begin{equation}
k(\phi)^{2} \, - \, \frac{(\sin^{4} \phi + \cos^{4} \phi)}{12} \, k(\phi)^{4}
\, + \, \gmathcal{O}\left(k(\phi)^{6}\right) \, = \, f_{\mu}             \label{second-order}
\end{equation}

\noindent
which gives

\begin{equation}
k(\phi)^{2} \, = \, f_{\mu} + \frac{(\sin^{4} \phi + \cos^{4} \phi)}{12} \,
f_{\mu}^{2} \, + \, \gmathcal{O}\left(f_{\mu}^{3}\right)  \,\, .            \label{order-f-3}
\end{equation}

Consider, for instance, the first of Eqs.~(\ref{B-order-0}), which is of the form
(\ref{typical-integral-0}) with $g(\gmathbf{k})=1$ and $\gamma = m_{2}^{(1)} - \mu^{(1)}$.
Equation (\ref{polar-integral} ) now becomes:

\begin{equation}
\delta \, = \, \frac{1}{8\pi^2} \int_{0}^{2\pi} \! d \phi \, k(\phi)^{2} \, \, .
                                                                      \label{delta=integtal}        
\end{equation}

\noindent
Introducing the notation $f_{\mu}$ as above, for small values of $\delta (>0)$ 
we obtain:

\begin{equation}
\delta = \frac{1}{4\pi} \left[ f_{\mu} + \frac{1}{16} f_{\mu}^{2}
+ \gmathcal{O} \left( f_{\mu}^{3} \right)  \right]                        \label{delta-vs-f1}
\end{equation}

\noindent
where use has been made of Eq.~(\ref{order-f-3}).
Inverting this relation we obtain eventually:

\begin{equation}
f_{\mu} \, = \, 4 \pi \delta \, - \, \pi^{2} \delta^{2} 
+ \gmathcal{O} \left( \delta^{3} \right)   \,\, ,                         \label{f1-vs-delta}   
\end{equation}

\noindent
which confirms the fact that $f_{\mu} = \gmathcal{O}(\delta)$.

In general, the case of a $\gmathbf{k}$-dependent $g$  can be treated by expanding 
$g(\gmathbf{k})$ about $|\gmathbf{k}|=0$, since the last integral on the rignt-hand 
side of Eq.~(\ref{polar-integral}) is restricted to $|\gmathbf{k}| < k(\phi) = 
\gmathcal{O}(\delta^{1/2})$, and  by retaining the relevant orders in $\delta$ in 
the final expression.
The integral over the {\it whole\/} BZ (i.e., the first term on the rignt-hand side of 
Eq.~(\ref{polar-integral})), on the other hand, can usually be performed analytically.

The Brillouin zone sums involving the Dirac delta functions or its derivatives
can be further evaluated according to the following device.
Writing
             
\begin{eqnarray}
\delta \left( T_{e}(\gmathbf{k}) + \gamma \right) & = &
\frac{\partial}{\partial\gamma} \, \Theta\left(T_{e}(\gmathbf{k})+\gamma\right)  \nonumber \\      
& = &  \frac{1}{\cos (Q/2)} \frac{\partial}{\partial f_{\mu}} \, 
\Theta \left( T_{e}(\gmathbf{k}) + \gamma \right)   \,\, ,  
                                                                    \label{smart-derivative}
\end{eqnarray} 

\noindent
we obtain for the Brillouin zone sum

\begin{eqnarray}
& \frac{1}{\gmathcal{N}} & \sum_{\gmathbf{k}}^{BZ} \, g(\gmathbf{k}) \, \delta^{(n)}
\left( T_{e}(\gmathbf{k}) + m_{2}^{(1)} - \mu^{(1)} \right)                \label{delta-n} \\   
& = & \frac{1}{(\cos (Q/2))^{(n+1)}} \, 
\frac{\partial^{(n+1)}}{\partial (f_{\mu}^{(1)})^{(n+1)}}
\frac{1}{\gmathcal{N}} \sum_{\gmathbf{k}}^{BZ} \, g(\gmathbf{k}) \, \Theta 
\left( T_{e}(\gmathbf{k}) + m_{2}^{(1)} - \mu^{(1)} \right)                  \nonumber \\    
& = & - \frac{1}{(\cos (Q/2))^{(n+1)}} \, 
\frac{\partial^{(n+1)}}{\partial (f_{\mu}^{(1)})^{(n+1)}} \,  
\frac{1}{4\pi^{2}}\int_{0}^{2\pi} \! d\phi \int_{O}^{k(\phi)} 
\! dk k \, g(k,\phi)                                                        \nonumber  
\end{eqnarray} 

\noindent
since the first term on the right-hand side of Eq.~(\ref{polar-integral}) does
not depend on $f_{\mu}$.

As an example, we evaluate the expression

\[
\frac{1}{\gmathcal{N}} \sum_{\gmathbf{k}}^{BZ} \, T_{o}^{2}(\gmathbf{k}) \,
\delta \left( T_{e}(\gmathbf{k}) + m_{2}^{(1)} - \mu^{(1)} \right)
\] 

\noindent
which appears in the first of Eqs.~(\ref{B-order-1}). 
According to Eq.~(\ref{smart-derivative}), we first evaluate the expression
(cf. also Eq.~(\ref{polar-integral})):                                                                   

\begin{eqnarray}
& \frac{1}{\gmathcal{N}} & \sum_{\gmathbf{k}}^{BZ} \, T_{o}^{2}(\gmathbf{k}) \, 
\Theta \left( T_{e}(\gmathbf{k}) + m_{2}^{(1)} - \mu^{(1)} \right)        \label{sum-To-2} \\
& = & 4 \, \sin ^{2}(Q/2) - \frac{1}{\pi^2} \sin ^{2}(Q/2)
\int_{0}^{2\pi}  d \phi \int_{0}^{k(\phi)} d k \, k \,
\left( \sin (k \cos \phi) + \sin (k \sin \phi) \right)^{2}         \,.    \nonumber
\end{eqnarray} 

\noindent
Expanding the factor within parentheses in the last integral in powers 
of $k$, as explained below Eq.~(\ref{f1-vs-delta}), we obtain:
 
\begin{eqnarray}
& \frac{1}{\gmathcal{N}} & \sum_{\gmathbf{k}}^{BZ} \, T_{o}^{2}(\gmathbf{k}) \, 
\Theta \left( T_{e}(\gmathbf{k}) + m_{2}^{(1)} - \mu^{(1)} \right)     \nonumber \\
& = & 4 \sin ^{2}(Q/2) - \frac{1}{\pi^{2}} \sin ^{2}(Q/2)
\int_{0}^{2\pi} \! d \phi \int_{0}^{k(\phi)} \! dk \, k \,
\left( k^{2} (\cos \phi+ \sin \phi)^{2} + 
\gmathcal{O}\left(k^{4}\right) \right)                                 \nonumber \\
& = & \sin ^{2}(Q/2) \left( 4- 8\pi \delta^{2} + \gmathcal{O}(\delta^{3}) \right) \, .
                                                                      \label{sum-To-2-again} 
\end{eqnarray}

\noindent
Using Eq.~(\ref{smart-derivative}) we obtain eventually:

\begin{eqnarray}      
& \frac{1}{\gmathcal{N}} & \sum_{\gmathbf{k}}^{BZ} \, T_{o}^{2}(\gmathbf{k}) \, \delta
\left( T_{e}(\gmathbf{k}) + m_{2}^{(1)} - \mu^{(1)} \right)     \label{sum-To-2-once-finally} \\
& = & \frac{1}{\cos (Q/2)} \frac{\partial}{\partial f_{\mu}} \,   
\frac{1}{\gmathcal{N}} \sum_{\gmathbf{k}}^{BZ} \, T_{o}^{2}(\gmathbf{k}) \, \Theta
\left( T_{e}(\gmathbf{k}) + m_{2}^{(1)} - \mu^{(1)} \right)            \nonumber  \\      
& = & - \frac{\sin ^{2}(Q/2)}{\cos (Q/2)} \, \frac{1}{\pi} \, f_{\mu} 
+ \gmathcal{O}\left( (f_{\mu}^{2} \right)          
\, = \, -4 \, \delta \, \frac{\sin ^{2}(Q/2)}{\cos (Q/2)}
+ \gmathcal{O}(\delta^{2}) \, .                                        \nonumber 
\end{eqnarray}

\noindent
With these prescriptions, the results (\ref{B-m}), (\ref{B-mm}), and (\ref{B-long})
are readily obtained.

It was shown in Section 3 that, as far as the matrix elements of the correlation 
function are concerned, only the explicit expressions of $d(\gmathbf{q},\omega)$ 
and $e(\gmathbf{q},\omega)$ are required at the order in $t/U$ we are considering
in this paper.
Writing

\begin{eqnarray}
d(\gmathbf{q},\omega) & = &  -1 - \tilde{\omega} + \alpha(\gmathbf{q}) + 
\gmathcal{O}\left(\left(t/U\right)^{3}\right)                       \nonumber \\
& = &  I(\gmathbf{q},\omega) - J(\gmathbf{q},\omega) - K(\gmathbf{q},\omega)
+ M(\gmathbf{q},\omega)                                                  \label{d-vs-I-J-K-M}          
\end{eqnarray}

\noindent
and

\begin{equation}
e(\gmathbf{q},\omega) \, = \, L(\gmathbf{q},\omega) + L(-\gmathbf{q},-\omega) + N(\gmathbf{q},\omega)  
                                                                            \label{e-vs-L-N}
\end{equation}

\noindent
where the quantities $I,J,\cdots$ are specified below, we obtain for the sums over the 
wave vector $\gmathbf{k}$ using the method described in this Appendix:

\begin{eqnarray}
I(\gmathbf{q},\omega) & \equiv & \frac{U}{\gmathcal{N}} \, \sum_{\gmathbf{k}}^{BZ} \,
\gmathcal{F}_{21}(\gmathbf{k},\gmathbf{k-q},\omega) \, = \,
\frac{-1+\delta}{1-\tilde{\omega}-\delta}                                   \nonumber \\
& + & \left(\frac{t}{U}\right) \, 2 \, \delta \, 
\cos (Q/2) \, (\cos q_{x}+\cos q_{y})                                          \label{I} \\
& + & \left(\frac{t}{U}\right)^{2} \, 4 \, \cos ^{2}(Q/2) \, (\cos q_{x}+\cos q_{y}) 
+ \gmathcal{O}\left(\left(t/U\right)^{3}\right)  \,\, ,                      \nonumber  
\end{eqnarray}

\begin{eqnarray}
J(\gmathbf{q},\omega) & \equiv & \frac{U}{\gmathcal{N}} \, \left(\frac{t}{U}\right)^{2} \,
\sum_{\gmathbf{k}}^{BZ} \, T_{o}^{2}(\gmathbf{k}) \,
\gmathcal{F}_{21}(\gmathbf{k},\gmathbf{k-q},\omega)                            \nonumber \\
& = & - 4 \left(\frac{t}{U}\right)^{2} \sin ^{2}(Q/2)  
+ \gmathcal{O}\left(\left(t/U\right)^{3}\right)  \,\, ,                             \label{J} 
\end{eqnarray}

\begin{eqnarray}
K(\gmathbf{q},\omega) & \equiv & \frac{U}{\gmathcal{N}} \, \left(\frac{t}{U}\right)^{2} \,
\sum_{\gmathbf{k}}^{BZ} \, T_{o}^{2}(\gmathbf{k}-\gmathbf{q}) \,
\gmathcal{F}_{21}(\gmathbf{k},\gmathbf{k-q},\omega)                            \nonumber \\
& = & - 4 \left(\frac{t}{U}\right)^{2} \sin ^{2}(Q/2)  
+ \gmathcal{O}\left(\left(t/U\right)^{3}\right) \,\, ,                              \label{K} 
\end{eqnarray} 

\begin{eqnarray}
L(\gmathbf{q}, \omega) & \equiv & \frac{U}{\gmathcal{N}} \, \left(\frac{t}{U}\right)^{2} \,
\sum_{\gmathbf{k}}^{BZ} \, T_{o}(\gmathbf{k}) \, T_{o}(\gmathbf{k}-\gmathbf{q}) \,
\gmathcal{F}_{21}(\gmathbf{k},\gmathbf{k-q},\omega)                            \nonumber \\
& = & - 2 \left(\frac{t}{U}\right)^{2} \sin ^{2}(Q/2) \, (\cos q_{x}+\cos q_{y}) 
+ \gmathcal{O}\left(\left(t/U\right)^{3}\right) \,\, ,                              \label{L}  
\end{eqnarray}

\begin{eqnarray}
M(\gmathbf{q},\omega) & \equiv & \frac{U}{\gmathcal{N}} \, \left(\frac{t}{U}\right)^{2} \, 
\sum_{\gmathbf{k}}^{BZ} \, T_{o}^{2}(\gmathbf{k}-\gmathbf{q}) \,
\gmathcal{F}_{22}(\gmathbf{k},\gmathbf{k-q},\omega)                            \nonumber \\
& = & - 2 \,\delta \, \left(\frac{t}{U}\right) \, 
\frac{\sin ^{2}(Q/2) \, (\sin q_{x}+\sin q_{y})^2}
{\cos (Q/2) \, (\cos q_{x}+\cos q_{y}-2)} 
+ \gmathcal{O}\left(\left(t/U\right)^{3}\right) \,\, ,                              \label{M}
\end{eqnarray}

\begin{eqnarray}
N(\gmathbf{q},\omega) & \equiv & \frac{U}{\gmathcal{N}} \, \left(\frac{t}{U}\right)^{2} \, 
\sum_{\gmathbf{k}}^{BZ} \, T_{o}(\gmathbf{k}) \, T_{o}(\gmathbf{k}-\gmathbf{q}) \,
\gmathcal{F}_{22}(\gmathbf{k},\gmathbf{k-q},\omega)                           \nonumber \\
& = & \left(\frac{t}{U}\right)^{2} \, \gmathcal{O}(\delta) 
= \gmathcal{O}\left(\left(t/U\right)^{3}\right) \,\, ,                              \label{N}
\end{eqnarray} 
where these results hold for $|\gmathbf{q}|\gg k_F$, as discussed in Section 3-B.

For the relevant matrix elements (\ref{d-vs-I-J-K-M}) and (\ref{e-vs-L-N})
we obtain eventually :
                                      
\begin{eqnarray}
d(\gmathbf{q},\omega) & = & - 1 - \tilde{\omega} + \left(\frac{t}{U}\right)^{2} \, 4 \, 
\cos ^{2}(Q/2) \, (\cos q_{x}+\cos q_{y}-2) + 
8 \left(\frac{t}{U}\right) ^{2} \sin ^{2}(Q/2)                             \nonumber \\  
& + &  2 \, \left(\frac{t}{U}\right) \, \delta \, \cos (Q/2) \,
(\cos q_{x}+\cos q_{y}-2)                                                  \nonumber \\     
& - &   2 \, \left(\frac{t}{U}\right) \, \delta \,
\frac{\sin ^{2}(Q/2) \, (\sin q_{x}+\sin q_{y})^{2}}
     {\cos (Q/2) \, (\cos q_{x}+\cos q_{y}-2)} \,   
+ \, \gmathcal{O}\left(\left(t/U\right)^{3}\right)                                  \label{d}
\end{eqnarray}

\noindent
and

\begin{equation}
e(\gmathbf{q},\omega) \, = \, - 4 \,\left(\frac{t}{U}\right)^{2} \, \sin ^{2}(Q/2) \, 
(\cos q_{x}+\cos q_{y}) \, + \, \gmathcal{O}\left(\left(t/U\right)^{3}\right) \,\, . \label{e}
\end{equation}


\section{Heisenberg spin waves for a spiral configuration}

In this Appendix, we give a simplified derivation of the spin-wave dispersion relation
for a two-dimensional Heisenberg antiferromagnet in the presence of a spiral
incommensurate magnetic ground state.
Although the form of this spectrum is well known,\cite{Keffer,Cooper-et-al} it is
worthed to give here a compact derivation in terms of the set of local spin 
quantization axes utilized in Section 3 and in Appendix A. 

We consider the Heisenberg Hamiltonian

\begin{equation}
H \, = \, \frac{1}{2} \sum_{i,j} \, J_{i,j} \, \gmathbf{S}_{i} \cdot
      \gmathbf{S}_{j}                                          \label{Heisenberg-Hamiltonian}
\end{equation}

\noindent
for spin $1/2$, where the sum over the lattice sites $i$ and $j$ extends in principle 
over distant neighbors.
We transform next the spin locally as follows:

\begin{equation}
S^{\alpha}_{i} \rightarrow \sum_{\beta} \, T_{\alpha\beta}^{(i)} \,
S^{\beta}_{i}                                                         \label{transformation}
\end{equation}

\noindent
where $T^{(i)}$ represents the $3 \times 3$ matrix given by the spin part of
(\ref{A-T-cos-sin}).
The Hamiltonian (\ref{Heisenberg-Hamiltonian}) then becomes:

\begin{equation}
H \, = \, \frac{1}{2} \sum_{i,j} \, J_{i,j} \, \sum_{\alpha,\beta} \,
{S}_{i}^{\alpha} \, T_{\alpha\beta}^{(ji)} \, {S}_{j}^{\beta}
                                                          \label{new-Heisenberg-Hamiltonian}
\end{equation}

\noindent
where $T^{(ji)}$ contains the difference $\theta_{j} - \theta_{i}$ in the place of
$\theta_{i}$ of Eq.~(\ref{A-T-cos-sin}).

Writing further

\begin{equation}
S_{i}^{x} \, = \, \frac{1}{2}   \left( S_{i}^{+} \, + \,  S_{i}^{-} \right)
\,\,\, , \,\,\,
S_{i}^{y} \, = \, \frac{1}{2 i} \left( S_{i}^{+} \, - \,  S_{i}^{-} \right)
\,\,\, ,                                                             \label{definition-S-pm}
\end{equation}

\noindent
and carrying out the standard Holstein-Primakoff transformation 

\begin{eqnarray}
S_{i}^{\pm} & \cong & \sqrt{2S} \, a_{i}^{\pm}     \nonumber \\
S_{i}^{z} & = & a_{i}^{\dagger} a_{i} \, - \, S \, = \, n_{i} \, - \, S          \label{H-P}
\end{eqnarray}

\noindent
at the leading order away from perfect alignment (and for large values of $S$),
the Hamiltonian (\ref{new-Heisenberg-Hamiltonian}) becomes:

\begin{eqnarray}
H & \cong &\frac{S^{2}}{2} \sum_{i,j} \, J_{i,j} \, \cos (\theta_{j} - \theta_{i})
\, - \, S \, \sum_{i,j} \, n_{i} \, J_{i,j} \, \cos (\theta_{j} - \theta_{i}) \nonumber \\
& + & \frac{S}{4} \, \sum_{i,j} \, J_{i,j} \, \left\{ (a^{\dagger}_{i} a_{j} +
a^{\dagger}_{j} a_{i}) [\cos (\theta_{j} - \theta_{i}) + 1]  \right.          \nonumber \\
& + & \left. (a^{\dagger}_{i} a^{\dagger}_{j} +
a_{i} a_{j}) [\cos (\theta_{j} - \theta_{i}) - 1] \right\}  \,\, .     \label{H-approximate}
\end{eqnarray}

We introduce at this point the lattice Fourier transform, specify the choice  
$\theta_{i}=\gmathbf{Q}\cdot\gmathbf{R}_{i}$, and write: 

\begin{equation}
\sum_{i,j} \, a_{i} b_{j} \, \cos \gmathbf{Q}\cdot(\gmathbf{R}_{j}-\gmathbf{R}_{i}) \,
J_{i,j} \, = \, \frac{1}{2} \, \sum_{\gmathbf{k}}^{BZ} \, a(\gmathbf{k}) b(-\gmathbf{k}) \,
\left( J(\gmathbf{k}-\gmathbf{Q}) \, + \, J(\gmathbf{k}+\gmathbf{Q}) \right) \,\, ,
                                                                             \label{arci-FT}
\end{equation}

\noindent
where $a_{i}$ and $b_{j}$ is an arbitrary pair of operators and where

\begin{equation}
J(\gmathbf{k}) \, = \, \sum_{\gmathbf{R}} \, \exp(-i\gmathbf{k}\cdot\gmathbf{R}) \,
J(\gmathbf{R})  \,\, .                                                           \label{J-FT}
\end{equation}

\noindent
With the notation

\begin{equation}
\alpha(\gmathbf{k}) \equiv - S \ J(\gmathbf{Q}) \, +\, \frac{S}{2} \left( J(\gmathbf{k})
\, + \, \frac{J(\gmathbf{k}+\gmathbf{Q}) \, + \, J(\gmathbf{k}-\gmathbf{Q})}{2} \right)
                                                                    \label{definition-alpha}
\end{equation}

\noindent
and

\begin{equation}
\beta(\gmathbf{k}) \equiv - \frac{S}{2} \left( J(\gmathbf{k})
\, - \, \frac{J(\gmathbf{k}+\gmathbf{Q}) \, + \, J(\gmathbf{k}-\gmathbf{Q})}{2} \right)
                                                                     \label{definition-beta}
\end{equation}

\noindent
for any given value of $\gmathbf{Q}$, the Hamiltonian (\ref{H-approximate}) reduces 
to the form:

\begin{eqnarray}
H & \cong &  \, \frac{\gmathcal{N} S^{2}}{2} \, J(\gmathbf{Q}) \, + \,
\sum_{\gmathbf{k}}^{BZ} \, \left[ \alpha(\gmathbf{k}) \, a^{\dagger}(\gmathbf{k})
a(\gmathbf{k}) \right.                                                    \nonumber \\
& + & \left. \frac{\beta(\gmathbf{k})}{2} 
\left( a^{\dagger}(\gmathbf{k}) a^{\dagger}(-\gmathbf{k}) 
\, + \, a(\gmathbf{k}) a(-\gmathbf{k}) \right) \right]  \,\, .          \label{Hamiltonian-FT}
\end{eqnarray}

The standard Bogoliubov transformation can be used at this point to diagonalize 
the Hamiltonian (\ref{Hamiltonian-FT}).
Writing

\begin{equation}
a(\gmathbf{k}) \, = \, u(\gmathbf{k}) \, b(\gmathbf{k}) \, + \, 
v(\gmathbf{k}) \, b^{\dagger}(-\gmathbf{k})                                 \label{Bogoliubov}
\end{equation}

\noindent
where $u(\gmathbf{k}) = \cosh \psi(\gmathbf{k})$ and 
$v(\gmathbf{k}) = \sinh \psi(\gmathbf{k})$ with

\begin{equation}
\tanh 2\psi(\gmathbf{k}) \, = \, - \, \frac{\beta(\gmathbf{k})}{\alpha(\gmathbf{k})}
\, \equiv \, \gamma(\gmathbf{k}) \,\, ,                                \label{definition-psi}
\end{equation}

\noindent
the Hamiltonian (\ref{Hamiltonian-FT}) becomes eventually:

\begin{equation}
H \, = \, E_{0} \,+ \, E_{1} \, + \, \sum_{\gmathbf{k}}^{BZ} \, \varepsilon(\gmathbf{k}) \,
b^{\dagger}(\gmathbf{k}) \, b(\gmathbf{k}) \,\, .                              \label{H-final}
\end{equation}

\noindent
In this expression:

\begin{equation}
E_{0} \, =  \, \frac{\gmathcal{N} S^{2}}{2} \, J(\gmathbf{Q}) \, ,  \label{definition-E-0}
\end{equation}

\begin{equation}
E_{1} \, = \,- \sum_{\gmathbf{k}} \, \left( \alpha(\gmathbf{k}) \, v^{2}(\gmathbf{k})
\, + \, u(\gmathbf{k}) \, v(\gmathbf{k}) \, \beta(\gmathbf{k}) \right) \, , \label{definition-E-1}
\end{equation}

\noindent
and

\begin{eqnarray}
\varepsilon(\gmathbf{k}) & = & \sqrt{\alpha^{2}(\gmathbf{k}) \, - \, \beta^{2}(\gmathbf{k})} \nonumber \\
& = & S\ \sqrt{\left( J(\gmathbf{Q}) - J(\gmathbf{k}) \right) \left( J(\gmathbf{Q}) -
\frac{J(\gmathbf{k}+\gmathbf{Q}) \, + \,
  J(\gmathbf{k}-\gmathbf{Q})}{2} \right)} \, .  
 \label{definition-epsilon}
\end{eqnarray}

\noindent
The spectrum (\ref{definition-epsilon}) coincides 
(apart for a different normalization of $J$) 
with that given in Ref.~\cite{Cooper-et-al} 
(cf. in particular Eq.~(67) therein).
Note from Eq.~(\ref{definition-epsilon}) the presence of a Goldstone mode for $\gmathbf{k}=0$ and
$\gmathbf{k}=\pm \gmathbf{Q}$ (owing to the symmetry $J(-\gmathbf{k})=J(\gmathbf{k})$).

In the classical limit (that is, for large values of $S$), minimization of the ground-state 
energy is equivalent to finding the minimum of $J(\gmathbf{Q})$. 
In this way, the appropriate value of $\gmathbf{Q}$ is determined.
For instance, when second ($J_{2}$) and third ($J_{3}$) nearest-neighbor couplings are 
included besides the nearest-neighbor 
antiferromagnetic
coupling ($J_{1}$), $J(\gmathbf{Q})$ reads:

\begin{equation}
J(\gmathbf{Q}) \, =  J_{1} \, (\cos Q_{x} + \cos Q_{y}) \, + \, 2 J_{2} \, \cos Q_{x} \cos Q_{y}
\, + \, J_{3} \, (\cos 2Q_{x} + \cos 2Q_{y}) 
\label{J-1-2-3}
\end{equation}

\noindent
(with the wave vectors measured in units of the inverse of the lattice spacing).
In this case, the diagonal spiral configuration is stable for $J_{3} > J_{1}/4 - J_{2}/2$
and $J_{3} > J_{2}/2$, implying that $J_{3}$ must be nonvanishing for the spiral phase
to be stable.\cite{Gelfand}
The corresponding spin-wave spectrum (\ref{definition-epsilon}) for
$J_{3} = J_{2} = J_{1}/5$ and $S=1/2$ is 
shown in Fig.~4 (full line), where the antiferromagnetic spectrum with $\gmathbf{Q}=(\pi,\pi)$ and 
$J_{3}=J_{2}=0$ (broken line) is also shown for comparison.

\begin{figure}[htb] 
\par \vspace*{.5cm} \par
   \centerline{
\psfig{file=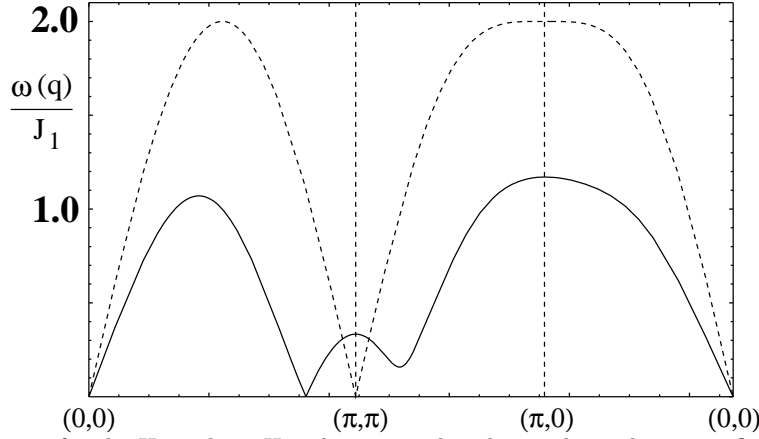,width=10.cm}
}
\caption{
Spin-wave spectrum for the Heisenberg Hamiltonian with a
diagonal-spiral spin configuration and couplings up to third nearest neighbors
specified in the text (full line).
The antiferromagnetic spectrum with nearest-neighbor coupling only is also
shown for comparison (broken line). 
}
\par \vspace*{.5cm} \par
 \end{figure}


\section{RKKY spin waves for a spiral configuration}

In this Appendix, we show how the part of the dispersion relation 
(\ref{disp-rel-4}) of the text containing the trigonometric functions in 
the denominator originates from the Ruderman-Kittel-Kasuya-Yosida (RKKY) 
interaction between two localized spins and mediated by the conduction 
electrons.\cite{Mattis,RK}
The long-range nature of this interaction is readily evidenced by expanding 
formally the denominator in Eq.~(\ref{disp-rel-4}) as a a power series in 
$\cos q_{x}$ and $\cos q_{y}$, so that an infinite number of Heisenberg-like 
terms appears.

We follow here the treatment by Mattis,\cite{Mattis} and consider
the interaction Hamiltonian between the localized spin operators $\gmathbf{S}_i$
(associated with the valence (filled) band) and the itinerant spin operators
$\gmathbf{s}_c(\gmathbf{R}_i)$ (associated with the conduction band), where the
suffix $i$ specifies the lattice site.
We write accordingly: 

\begin{equation}
H_{\gmathrm{exc}} \, = \, - I \, \sum_{i} \, \gmathbf{S}_i \cdot \gmathbf{s}_c(\gmathbf{R}_i)
                                                                               \label{H-exc}
\end{equation}

\noindent
where $I$ is an exchange integral.

For a generic spin-$1/2$ operator at site $i$ associated with the conduction
electrons we write: 

\begin{equation}
{s}_{i}^{\mu } \, = \, \frac{1}{2\gmathcal{N}} \sum_{\gmathbf{k},\gmathbf{q}} 
\sum_{\xi,\xi'} \, \gmathrm{e}^{-i \gmathbf{q} \cdot \gmathbf{R}_{i}}  \, 
d_{\gmathbf{k}+\gmathbf{q} \xi}^{\dagger} \, \sigma^{\mu}_{\xi,\xi'} \,   
d_{\gmathbf{k} \xi'}                                                     \label{s-conduction}
\end{equation}

\noindent
where $\mu=(+,-,z)$ and $d_{i \xi}$ are the destruction operators along the
local spin-quantization axes.
Combining these operators with the eigenvectors of the mean-field Hubbard Hamiltonian 
[cf. Eq.~(\ref{A-gamma-definition})] and recalling the notation (\ref{A-F-definition}),
we cast Eq.~(\ref{s-conduction}) in the form:
                                                                   
\begin{equation}
{s}_i^{\mu } \, =  \, \frac{1}{2\gmathcal{N}} \sum_{\gmathbf{k},\gmathbf{q}} \sum_{r, r'}
\, \gmathrm{e}^{-i \gmathbf{q} \cdot \gmathbf{R}_{i}} \, 
\gamma_{\gmathbf{k}+\gmathbf{q},r}^{\dagger} \, 
F^{\mu}_{r,r'}(\gmathbf{k}+\gmathbf{q},\gmathbf{k}) \, \gamma_{\gmathbf{k},r'} \, \, .
                                                                        \label{spin-compact}
\end{equation}

\noindent
The restriction to the conduction band implies that $r=r'=2$ in 
Eq.~(\ref{spin-compact}) when $\delta >0$.
In this case the Hamiltonian (\ref{H-exc}) becomes:

\begin{equation}
H_{\gmathrm{exc}} \, = \, - \frac{I}{2\gmathcal{N}} \, \sum_{i} \sum_{\mu} \, 
\left(S_{i}^{\mu}\right)^{\dagger} \, \left( \sum_{\gmathbf{k},\gmathbf{q}} \,
e^{-i \gmathbf{q} \cdot \gmathbf{R}_{i}} \, \gamma_{\gmathbf{k}+\gmathbf{q},2}^{\dagger} \, 
F^{\mu}_{22}(\gmathbf{k}+\gmathbf{q},\gmathbf{k}) \, \gamma_{\gmathbf{k},2} \right)
\,\, .                                                                                                                     
                                                                         \label{H-exc-final}
\end{equation}                               

Applying at this point the standard procedure \cite{Mattis} to obtain the energy shift
of second-order in the Hamiltonian (\ref{H-exc-final}), the following RKKY-type effective 
Hamiltonian results:

\begin{equation}
H^{RKKY} \, =  \, \sum_{i j} \sum_{\mu \nu}  \, \left(S_{i}^{\mu}\right)^{\dagger}  
                \, J_{i j}^{\mu \nu} \, S_{j}^{\nu}\                          \label{H-RKKY}
\end{equation}

\noindent
with the notation

\begin{equation}
J_{i j}^{\mu \nu} \, = \, \frac{I^{2}}{4\gmathcal{N}^{2}} \, \sum_{k<k_{F}} 
\sum_{|\gmathbf{k}+\gmathbf{q}|> k_{F}} \,
\gmathrm{e}^{-i \gmathbf{q} \cdot (\gmathbf{R}_{i}-\gmathbf{R}_{j})} \,
\frac{ F^{\mu}_{2 2}(\gmathbf{k}+\gmathbf{q},\gmathbf{k}) \, 
F^{\nu *}_{2 2}(\gmathbf{k}+\gmathbf{q},\gmathbf{k})}
{\epsilon_{2}(\gmathbf{k}) \, - \, \epsilon_{2}(\gmathbf{k}+\gmathbf{q})} \, \, . 
                                                                              \label{J-RKKY}
\end{equation}
             
\noindent
Note that the expression (\ref{J-RKKY}) does not vanish even when sites
$i$ and $j$ are far apart.
Note also that the energy denominator on the right-hand side never vanish
by construction.

There remains to obtain the spin-wave dispersion associated with the spin
Hamiltonian (\ref{H-RKKY}).
To this end, we introduce in Eq.~(\ref{H-RKKY}) the usual Holstein-Primakoff 
transformation for the spin operators and truncate the expansion in the
bosonic operators $a$ and $a^{\dagger}$ to quadratic order 
[cf. Eq.~(\ref{H-P})], to obtain:

\begin{eqnarray}
H^{RKKY} & = & E_{0} \, + \, S \, \sum_{|\gmathbf{q}| \gg k_{F}} 
\left[ \left( J^{++}(\gmathbf{q}) + J^{--}(-\gmathbf{q}) \right) 
a^{\dagger}(\gmathbf{q}) \, a(\gmathbf{q}) \right.               \nonumber \\
& + & \left. J^{+-}(\gmathbf{q}) \, a^{\dagger}(\gmathbf{q}) \, a^{\dagger}(-\gmathbf{q})
\, + \, J^{-+}(-\gmathbf{q}) \, a(\gmathbf{q}) \, a(-\gmathbf{q}) \right]
                                                                           \label{H-RKKY-HP}
\end{eqnarray}

\noindent                                                               
where now 

\begin{equation}
E_{0} \, = \, S^{2} \, \sum_{|\gmathbf{q}| \gg k_{F}} J^{zz}(\gmathbf{q}) \,
+ \, S \, \sum_{|\gmathbf{q}| \gg k_{F}} J^{--}(\gmathbf{q})     \label{def-E0}
\end{equation}

\noindent
with the restriction $|\gmathbf{q}| \gg k_{F}$ replacing the weaker condition
$|\gmathbf{k}+\gmathbf{q}| > k_{F}$ (cf. Section 3-B).
Note that in Eq.~(\ref{H-RKKY-HP}) we have introduced the notation

\begin{equation}
J^{\mu \nu}(\gmathbf{q}) \, = \, \frac{I^{2}}{4\gmathcal{N}} \sum_{|\gmathbf{k}|<k_{F}}  
\frac{ F^{\mu}_{2 2}(\gmathbf{k}+\gmathbf{q},\gmathbf{k}) \,
       F^{\nu*}_{2 2}(\gmathbf{k}+\gmathbf{q},\gmathbf{k})}
     {\epsilon_{2}(\gmathbf{k}) \, - \, \epsilon_{2}(\gmathbf{k}+\gmathbf{q})}  
                                                                             \label{J-mu-nu}
\end{equation} 

\noindent            
and used the identity $\sum_{i} J_{ij}^{\mu \nu}=0$ for any given $j$.

The quadratic Hamiltonian (\ref{H-RKKY-HP}) is then diagonalized by the 
standard Bogoliubov transformation (cf. Appendix D).
The resulting spin-wave spectrum has the form: 

\begin{equation}
\omega^{RKKY}(\gmathbf{q}) \, = \, S \left[ \left( J^{++}(\gmathbf{q}) + 
J^{--}(-\gmathbf{q}) \right)^{2} \, - \, 4 \, J^{+-}(\gmathbf{q}) \,
J^{-+}(-\gmathbf{q}) \right]^{1/2} \, \, .                                 \label{omega-RKKY}
\end{equation}

To proceed further, we need to specify the quantities $F^{+}_{22}$, $F^{-}_{22}$,
and $\epsilon_{2}$ entering the definition (\ref{J-mu-nu}), at the relevant
order in the small parameter $t/U$.
Using Eqs.~(\ref{epsilon-2})-(\ref{W2}) we obtain for $\delta>0$: 

\begin{eqnarray}
F^{+}_{22}(\gmathbf{k}+\gmathbf{q},\gmathbf{k}) & = & i \sqrt{2} \,
\left(\frac{t}{U}\right) \, T_{o}(\gmathbf{k}+\gmathbf{q}) \, + \,
\gmathcal{O}\left( (t/U)^{2} \right)                      \nonumber \\
F^{-}_{22}(\gmathbf{k}+\gmathbf{q},\gmathbf{k}) & = & - \, i \sqrt{2} \,
\left(\frac{t}{U}\right) \, T_{o}(\gmathbf{k}) \, + \,
\gmathcal{O}\left( (t/U)^{2} \right)                                    \label{F-approximate}
\end{eqnarray}

\noindent
as well as
 
\begin{equation}
\epsilon_{2}(\gmathbf{k}) \, - \, \epsilon_{2}(\gmathbf{k}+\gmathbf{q}) \,
= \, t \, \left( T_{e}(\gmathbf{k}) \, - \, T_{o}(\gmathbf{k}+\gmathbf{q})
\right) \, + \, \gmathcal{O}\left( (t/U)^{2} \right) \,\, ,         \label{epsil-epsil-appox}
\end{equation}

\noindent
where we have approximated $2m_{2}^{(0)} = 1-\delta \cong 1$.

The method developed in Appendix C can be used at this point to perform 
the sum over $\gmathbf{k}$ in Eq.~(\ref{J-mu-nu}).
The result is:

\begin{equation}
J^{++}(\gmathbf{q})\, \cong \, - \, I^{2} \, \frac{t}{U^{2}} \, \delta \,
\frac{\sin^{2} (Q/2)}{\cos (Q/2)} \,
\frac{(\sin q_{x} + \sin q_{y})^{2}}{\cos q_{x} + \cos q_{y} - 2} \, ,
                                                                           \label{J++approx}
\end{equation}

\begin{equation}
J^{--}(\gmathbf{q}) \, \cong \, - \, I^{2} \, \frac{t}{U^{2}} \,
\gmathcal{O}(\delta^{2}) \, ,                                               \label{J--approx}
\end{equation}

\begin{equation}
J^{+-}(\gmathbf{q}) \, = \, J^{-+}(\gmathbf{q}) \, \cong \, I^{2} \, 
\frac{t}{U^{2}} \, \gmathcal{O}(\delta^{2}) \, .                            \label{J+-approx}
\end{equation}

\noindent
At the lowest order, only $J^{++}$ contributes to Eq.~(\ref{omega-RKKY}), 
which then reduces to:

\begin{equation} 
\omega^{RKKY}(\gmathbf{q}) \, \cong \, S \, I^{2} \, \frac{t}{U^{2}} \, \delta 
\, \frac{\sin^{2} (Q/2) \, (\sin q_{x} + \sin q_{y})^{2}}
        {|\cos (Q/2)| \, (2 - \cos q_{x} - \cos q_{y})} \,\, .
                                                                   \label{omega-RKKY-approx}
\end{equation}

The self-consistency equation (\ref{cos}) relating $Q$ to $\delta$ and $t/U$
for the diagonal spiral configuration can eventually be used, to yield (for
$S=1/2)$:
                                              
\begin{equation}
\omega^{RKKY}(\gmathbf{q}) \, \cong \, J_{\gmathrm{eff}} \, 
\frac{(\sin q_{x} + \sin q_{y})^{2}}{2 - \cos q_{x} - \cos q_{y}}
                                                             \label{omega-RKKY-approx-final}
\end{equation}

\noindent
with $J_{\gmathrm{eff}}$ given by Eq.~(\ref{delta-c}) of the text and where we have 
set $I=2U$.
Note that Eq.~(\ref{omega-RKKY-approx-final}) coincides with Eq.~(\ref{RKKY})
of the text, which was obtained by setting ``by hand'' 
$\gmathcal{F}_{21}=\gmathcal{F}_{12}=0$ in the full calculation.
Note also that the expression (\ref{omega-RKKY-approx-final}) involves transverse
spin components only, akin the general expression (\ref{omega-2}) of the text.





\begin{thebibliography}{99}
\bibitem{footnote-1}
It can be further verified that $\omega(\gmathbf{q})$ given by Eq.~(\ref{disp-rel-4}) 
is symmetric under the separate transformations
$(q_{x} \rightarrow q_{y}, q_{y} \rightarrow q_{x})$ and 
$\gmathbf{q} \rightarrow - \, \gmathbf{q}$.
\end{thebibliography}
\end{document}